\documentclass[prd,a4paper,nofootinbib,
showpacs, 
]{revtex4}
\usepackage{axodraw}
\def\eq#1{{eq.~(\ref{#1})}}
\def\eqs#1#2{{eqs.~(\ref{#1})--(\ref{#2})}}

\def\sec#1{{sec.~(\ref{#1})}}

\def\tab#1{{tab.~(\ref{#1})}}
\def\tabs#1#2{{tabs.~(\ref{#1})--(\ref{#2})}}
\def\vev#1{\left\langle #1\right\rangle}

\def\Tr{\mbox{Tr}\,}

\def\gtap{\ \raisebox{-.4ex}{\rlap{$\sim$}} \raisebox{.4ex}{$>$}\ }

\def\hbar{\hspace{0pt}\raisebox{1pt}{$-$} \hspace{-7pt} h}

\def\5{\overline 5}

\newcommand{\be}{\begin{equation}}
\newcommand{\ee}{\end{equation}}
\newcommand{\bea}{\begin{eqnarray}}
\newcommand{\eea}{\end{eqnarray}}
\newcommand{\nn}{\nonumber}

\begin{document}
\title[Flavor and electroweak symmetry breaking]{Flavor and electroweak symmetry breaking at the TeV scale}
\date{\today
}
\author{F.~Bazzocchi}
\author{M.~Fabbrichesi}
\affiliation{INFN, Sezione di Trieste and\\
Scuola Internazionale Superiore di Studi Avanzati\\
via Beirut 4, I-34014 Trieste, Italy}
\begin{abstract}

\noindent We present a unified picture of flavor and electroweak  symmetry breaking at the TeV scale. Flavor and Higgs bosons arise as  pseudo-Goldstone modes in a nonlinear sigma model. Explicit collective symmetry  breaking yields stable vacuum expectation values and masses protected at one loop by the little-Higgs mechanism.  The coupling to the fermions through  a Yukawa lagrangian with a $U(1)$ global flavor symmetry generates well-definite mass textures that correctly reproduce the mass hierarchies and mixings of  quarks and leptons. The model is more constrained than usual little-Higgs models because of bounds on weak and flavor physics. The main experimental signatures testable at the LHC are a rather large mass $m_{h^0} = 317\pm 80$ GeV for the (lightest) Higgs  boson   and a characteristic spectrum of new  bosons and fermions with masses around the TeV scale.
\end{abstract}
\maketitle
%
\vskip1.5em
\section{Motivations and background} 
\label{sec:mb}

The requirement of naturalness for the standard model with a light Higgs boson seems to demand new physics at or around the 1 TeV scale. On the other hand, precision measurements do not show any departure from standard physics up to roughly 10 TeV. This \textit{little hierarchy} problem can be solved in little-Higgs models~\cite{littlest,skiba,littlehiggs} by introducing new particles near 1 TeV, the effect of which is however sufficiently hidden not to show in the precision tests. 
 
 The Higgs sector of the standard model  also contains the physics of flavor, that is, the hierarchy in the fermion masses and the mixing among different generations. What happens if we try to include flavor symmetry and its  breaking in the little-Higgs models? At first sight, this seems impossible because of the much higher scale of the order of $10^4$ TeV at which flavor symmetry breaking should take place if we want it to agree with the known bounds on flavor changing neutral currents. However, these bounds depend on the specific realization of the symmetry breaking and they are not necessarily so strong if the flavor symmetry is only global and there are no flavor charged gauge bosons~\cite{federica}.
 
 As we shall show, it is possible to take  closely related breaking scales   for both the electroweak and flavor symmetries and thus unify the two into a single little-Higgs model. 
  Thanks to this unification we are able to both  solve the little hierarchy problem and provide  stable  textures in the mass matrices of a well-defined type that correctly reproduce the mass hierarchies and mixings of  quarks and leptons. The model has a characteristic spectrum testable at the LHC of new particles, in addition to those of the standard model, and  a lightest Higgs boson mass  more constrained than in the usual, electroweak only, little-Higgs models that turns out to be heavy, in the sense of preferring values larger than 200 GeV. 
  
  Some of the work in this paper has been already presented in Letter form~\cite{bf}. This paper includes additional results and a more detailed discussion of the model and its consequences.
   
\subsection{A stable standard model with a natural cut off around 10 TeV}

For the standard model to be valid up to a scale  around 10 TeV (as indicated by precision measurements), the amount of fine-tuning of the value of the bare mass required in order to keep the mass of the Higgs boson to its value of about a hundred GeV is of the order of 1\% for 1-loop (quadratically divergent) radiative corrections coming from top-quark loops:
\be
 \frac{\lambda_t \Lambda^2}{(4\pi)^2} \simeq (1 \; \mbox{TeV})^2\, ,
\ee
where $\Lambda = 4\pi f^2$ is the effective cut off and $f \simeq 1$ TeV the scale of the electroweak symmetry breaking, and 10\% for similar loop corrections coming from the gauge bosons and the Higgs itself:
\be
 \frac{g^2 \Lambda^2}{(4\pi)^2}  \simeq \frac{\lambda^2 \Lambda^2}{(4\pi)^2}   \simeq (700 \; \mbox{GeV})^2 \, .
\ee

This fine-tuning (the little hierarchy  problem)  can be considered a hint to new physics taking place at or around the 1 TeV scale. For instance, the problem would be greatly ameliorated if additional degrees of freedom were to cancel the 1-loop divergence and the first contribution  were to arise only  at the 2-loop order, thus providing an additional factor of roughly $(4\pi)^2$ in the suppression of these corrections and therefore bringing the amount of fine-tuning to an acceptable level.

This idea has provided the motivation for recent work on the so-called \textit{little Higgs models}~\cite{littlest,skiba,littlehiggs}. In these models the Higgs boson is first thought as a Goldstone boson and therefore exactly massless and with no potential. Its effective potential, and therefore mass and vacuum expectation value, is generated by an explicit symmetry breaking that however can only proceed by simultaneously breaking more than one symmetry. Because of this \textit{collective} symmetry breaking, the effective potential does not contain quadratically divergent mass terms---in particular from the gauge and heavy top quark loops---and therefore the value of the Higgs mass is made stable against 1-loop radiative corrections with very little tuning of the value of the bare mass.

Many examples of little Higgs models have been introduced and discussed in the literature~\cite{littlest,skiba,littlehiggs} and some of their experimental signatures (and constraints) have been already extensively discussed~\cite{littlehiggs2}. Even though some of these constraints are rather strong and tend to reintroduce some amount of fine tuning, we shall not be overly concerned  with their impact since we are looking more for a consistent framework than with its detailed realization (see, however, ref.~\cite{Cheng} for a simple way to lessen the constraints). In what follows, we are particularly interested  in the so-called \textit{littlest} Higgs model~\cite{littlest}, in which the Higgs boson is a pseudo-Goldstone boson of the spontaneous breaking of   a $SU(5)$ symmetry donw to $SO(5)$; at low energies, the model contains only one Higgs doublet. The littlest Higgs model is of relevance in what follows
because---as we shall argue in section \ref{sec:tfl}---it is embedded in the flavorless limit of the model presented here.

\subsection{Flavons and textures}

One of the most tantalizing clue for physics beyond the standard model that we know of comes from the Higgs sector and the closely related flavor structure.  Data on particle masses and mixing angles present us with a wealth of information not too dissimilar to that once offered by  Mendeleev's period table and seem to beg for a dynamical explanation of their regularities. These data are encoded in the standard model into the Yukawa lagrangian which gives mass to the fermions and shapes  their mixing and mass hierarchies. This lagrangian is thus controlled by a (large) number of parameters that appear to be arbitrary insofar as their values are chosen by hand to match the experimental data; moreover, their values must  be chosen in a precise manner and many of them vary across several orders of magnitude. The stability against radiative corrections of these patterns and hierarchies seems to require some amount of fine tuning. While any fine tuning of the parameters can always be seen  as a mere coincidence (or, perhaps more speculatively, as anthropic selection at work), we take here the point of view that its explanation---like that for the little hierarchy---calls for new physics.

 A first step in the direction of improving our understanding of the flavor structure of the standard model can be taken by re-organizing the parameters and considering the  mass matrices of quarks and leptons not as $3\times 3$ arbitrary matrices but as matrices having well-defined textures controlled by one or at most few parameters. In this picture, the mass matrices have entries that are powers of these few parameters modulated by dimensionless coefficients  of order one---and thus requiring no further explanation. While this is not yet a dynamical model---it is really just ``kinematics''---it helps in providing a framework in which to bring the dynamics eventually.

At least part of this dynamics comes from identifying the small parameters, the powers of which give rise to the textures, with the vacuum expectation values of some scalar fields with quantum numbers running across the horizontal family structure of fermions. In this picture---usually referred to as the Froggart-Nielsen mechanism~\cite{flavormodels-old}---the emerging textures are due to different charge assignments for the fermions, and therefore from the different powers of the small parameters, associated to the vacua of the scalar fields, making up the mass matrices arising from the Yukawa lagrangian.

The next, and crucial step consists in providing stability against radiative corrections for the patterns thus generated and therefore explaining away the apparent fine tuning of parameters encountered. This problem can be rephrased in terms of the naturalness of the dimensionful parameters of the model that must be protected against corrections that tends to bring all of them to the highest mass scale in the problem, usually the cut off of the effective field theory (all dimensionless parameters are then assumed of order 1 and therefore natural). Such a naturalness---in the 't Hooft's sense---is achieved by identifying the one or more  symmetries that would be recovered in the limit of vanishing interactions and vacuum expectation values.

This problem of fine tuning and overall stability is present in all models trying to describe the mass matrices of the fermions, and the textures by which they are characterized, in terms of the vacuum expectation value $v$ of one, or more,  scalar fields, the flavons~\cite{flavormodels-old,flavormodels-new}. In these models, the texture is  written in terms of the ratio $\varepsilon = v/f$, where $f$ is now the  flavor symmetry breaking scale and
$\varepsilon$ is the small parameter of the texture typically of the order of the Cabibbo angle. These patterns tend however to be washed out by the quadratically divergent radiative corrections to the mass term $\mu^2$ that make the vacuum expectation value, for a generic quartic potential proportional to a parameter $\lambda \simeq 1$,
\be
v^2 \simeq \mu^2 \simeq f^2
\ee
and therefore  $\varepsilon =1$.

A small $\varepsilon$ comes in a natural manner if the mass term is protected at the one loop level and only logarithmically divergent so that
\be
v^2 = \mu^2 \simeq \frac{\log (\Lambda^2/f^2)}{(4 \pi)^2}\, f^2
\ee
and we have $\varepsilon^2 \ll 1$ independently of the scale $f$.
For this reason, the same collective symmetry breaking mechanism as in the little Higgs mechanism and the  identification of the flavons as pseudo-Goldstone bosons of some horizontal symmetry were successfully applied in a recent series of papers~\cite{littleflavons,federica} to the problem of flavor physics.

In this approach---referred to, for obvious reasons, as the \textit{little flavon model}---an $SU(6)$ symmetry is spontaneously broken to $Sp(6)$ thus giving rise to 14 (real) Goldstone bosons (a similar model has been discussed before in the context of weak physics in ref.~\cite{skiba}). Four subgroups $[SU(2)\times U(1)]^2$ are gauged and their gauge couplings explicitly break the global $SU(6)$ symmetry thus giving rise to a potential for the, at this point, two pseudo-Goldstone bosons that are doublets under the flavor symmetry. The quadratically divergent contribution to the 1-loop potential are however forbidden by the little-Higgs mechanism of collective breaking for which only logarithmic divergencies are allowed.

The surviving flavor symmetry is $SU(2)\times U(1)$, the spontaneous symmetry breaking of which comes from the Yukawa coupling to the right-handed neutrinos. The Yukawa lagrangian couples the little flavons to the fermions, accordingly with the charge assignment chosen, and, after spontaneous symmetry breaking, gives raise to the mass matrices with the desired textures in terms of the vacuum expectation values of the two flavor doublets. It is shown in~\cite{littleflavons} that such a model gives rise to characteristic textures that correctly reproduces all fermion masses and mixing angles.

\subsection{Little Higgs meets little flavon}  

As viable as the little flavon model is, it leaves open the question of how the flavons  and the weak Higgs field can be accomodated within an unified picture. Since the mass textures are but a modulation of the vacuum expectation value of the Higgs boson, the interplay between electroweak and flavor symmetries must lead us toward an unified picture of the two symmetries and their spontaneous breaking at closely related, if not the same, scales. 

The energy scale of any horizontal flavor symmetry breaking is usually thought as well separated from that of the electroweak symmetry breaking mainly because of the constraints on neutral flavor changing currents. The experimental bounds on flavor changing processes set rather stringent constraints on the value of the scale $f$ at which flavor symmetry must be broken. In the case of the little flavon model, this scale turns out~\cite{federica} to be between $10^{3}$ and $10^4$ TeV. This bound comes from processes---like $K^0$-$\bar K^0$ mixing---mediated by the flavor gauge bosons. It puts the flavor scale several order of magnitude higher than that of electroweak physics and makes it difficult to think of them in a unified manner. Moreover, radiative corrections from the flavor  to the electroweak sector become dangerously large and bring back into the picture some unwelcome fine tuning. However, this result is heavily based on assuming the horizontal symmetry to be local and therefore having to include the effect of the corresponding  gauge  bosons. In the absence of these, the constraint can be relaxed and, depending on the specific model, the energy scale made closer and even the same as  that of electroweak physics.

Interesting enough, contrary to those for the flavor gauge bosons, direct bounds on the effect of the scalar flavons are not very restrictive, giving, at least for some specific model, a scale  $f$ of the order of the TeV. This observation suggests to make the flavor symmetry into a global~\footnote{This global symmetry, as well as those of the little Higgs model,  must be thought as arising at same intermediate scale, well below that of string theory where all symmetries are necessarily local.} (rather than local) symmetry and thus avoid the more stringent bounds on the gauge flavor bosons (that do not exist any longer) and  bring the flavor symmetry breaking scale closer to that of electroweak physics.
 The unification of flavor and electroweak symmetry is thus made possible and an explicit example of it is the main result of this work.

Needless to say, the above scenario---that is going to be realized in the model that follows---is still far from being a complete theory of flavor. In particular, it leaves open the question of the absolute value of the fermion masses, most notably the large difference between those of  neutrinos and heavy quarks; this problem, and the much larger hierarchy implied, clearly requires a much deeper understanding of the dynamics in the ultraviolet and beyond the cut off of the model. Nevertheless,  the model we discuss does set the scene for a more profound dynamical understanding of the physics of flavor by creating a framework (with a special, and rather restrictive, choice of textures for the mass matrices of the fermions)  and identifying the relevant symmetries and degrees of freedom  at or around the  TeV scale that make an unification between flavor and electroweak physics possible within a natural model. In doing so, it says something specific about  physics in  the range to be explored by LHC, giving a (lightest) Higgs boson mass in a well defined range and particles in addition to those of the standard model to be discovered.

\section{The flhiggs model~\cite{bf}}  
\label{sec:tfm}

In order to have a  single, unified model \textit{\`a la} little Higgs describing  the entire flavor structure as well as the electroweak symmetry breaking,  the Higgs boson and the flavons must be  the pseudo-Goldstone bosons of the same spontaneously broken global symmetry.
These pseudo-Goldstone bosons---we shall call them \textit{flhiggs}---should transform under both  flavor  and    electroweak symmetries.
 The symmetry breaking should leave the electroweak  (or an extended symmetry, a subgroup of which is the electroweak) and the flavor  symmetry unbroken. In a further step,   the flavor and the electroweak symmetries  break leaving, as in the standard model,  the    electric charge $U(1)_Q$   as the only unbroken symmetry.

To construct such a model, it is necessary first to identify the
flavor and electroweak symmetry subgroups at the scale $f$ of the
spontaneous symmetry breaking of the global symmetry. The simplest
choice would seem to be a product of the flavor symmetry $G_F$ and
the electroweak symmetry $[SU(2)\times U(1)]_W$, where for the
flavor group $G_F$ we can take, without loss of generality,
$U(N)_F$. In this case the  flhiggs bosons  should transform, for
example if we take $U(N)_F$ to be $U(2) \simeq SU(2)\times U(1)$,
as doublets in the fundamental representations of the two $SU(2)$
groups, the electroweak and the $SU(2)$ in $U(2)_F$. However we
have to reject this choice since with the scalar fields as
doublets in both groups  the scale of flavor breaking will necessarily
coincide with that of the electroweak breaking with undesirable
 consequences for the phenomenology of the model.

 This holds true for
any choice of $G_F$. We are therefore necessarily lead to
extend the electroweak symmetry and the minimal extention gives us
$[U(N)]_F\times[SU(3)\times U(1)]_W$. In  this case if the flhiggs
bosons  transform in the fundamental representations of both
groups,  the breaking of the flavor symmetry can happen at a scale
different from the electroweak, that is, there is a limit in which
the flavor symmetry is broken  and its breaking induces the
breaking of the $[SU(3)\times U(1)]_W$ electroweak symmetry to the
standard $[SU(2)\times U(1)]_W$. This extention brings into the
model an extra neutral gauge boson and exotic fermion states necessary to complete the weak
doublets. These additional states give rise to new physics with crucial phenomenological consequences for the model. The mass and mixing of the extra gauge boson affect the neutral currents and impose rather severe bounds on the  parameters of the model. Moreover, the masses and mixing of the exotic  with standard fermions must be controlled by some additional symmetry that we take for simplicity to be an ableian $U(1)$. The exotic fermions, being charged under this abelian symmetry, only weakly couple to the standard fermions and acquire heavier masses.

\subsection{What is the horizontal flavor symmetry?}

The flavor symmetry could, in principle be abelian or nonabelian, that is a $U(2)$ since, for three generation at least, a $U(3)$ would introduce no differentiation.
Let then consider the nonabelian $SU(2)$ case.~\footnote{Which is the symmetry discussed in the little flavon model of ref.~\cite{littleflavons}.}
 The   flhiggs bosons arising as pseudo-Goldstone bosons in a model of this kind are in the fundamental representations of both the flavor and the weak $SU(2)$ groups. Therefore, they transform as $(2,3)$ and $(2,\overline{3})$ under the flavor-electroweak symmetry. 
 
 In order to construct flavor-electroweak invariant Yukawa term we have to choose the representations for the standard fermions. The left-handed fermions have to transform as a $3$ of $[SU(3)\times U(1)]_W$ since we want a doublet when $[SU(3)\times U(1)]_W$ is broken to $[SU(2)\times U(1)]_W$. As already noticed, we are obliged to introduce at least one exotic left-handed fields for each quark and lepton family. On the contrary, the right-handed ones could be singlets of $SU(3)$ or the third component of a triplet (anti-triplet) of weak $SU(3)$. Notice that in the latter case we would have to introduce other two exotic right-handed fermions for each quark and lepton family. 
 
 We still have  to assign the representations with respect to the flavor group. We could have singlets, doublets or triplets. We reject  the last case since it is impossible to reproduce the right hierarchies  by this choice for either left-handed or right-handed fermions or both. If we use singlets and doublets we have, for instance, that  two left-handed fermion families form a flavor doublet and the third is a singlet. In this case the choice for the left-handed fermion representations severely restricts that of the right-handeds  while there is no mixing between the doublet and the singlet.

Consider for example the Yukawa term for the charged leptons and
suppose that the second and the third family are in a flavor
doublet, while the first family is a singlet  with respect to the
flavor symmetry. This assignment  is motivated by the results obtained in \cite{littleflavons}. Each left-handed lepton family forms a triplet of
$SU(3)_W$, so we have
 \bea
 \label{example}
 L^e_L=\Bigg (\begin{array}{c}
   \nu^{e}_{L}\\
   e_{L}\\
    \tilde{e}_{L}\\
 \end{array}\Bigg)=(1,3)_L& \quad &
 E_L=\Bigg (\begin{array}{c}
   \nu^{i}_{L}\\
   e^{i}_{L}\\
    \tilde{e}^{i}_{L}\\
 \end{array}\Bigg)=(2,3)_L\,.
\eea
with $i=1,2$ , $e^1=\mu$ and $e^2=\tau$ and an exotic lepton for each family. In \eq{example1} we have indicated  in the brackets the fields representations with respect to flavor $SU(2)$ and weak $SU(3)$, respectively.
There are only
two possible choices for the representations of the right-handed
 charged leptons in order to have a Yukawa term involving $L^e_L$ that give mass to the electron and these are
  \bea
\label{example1}
  \overline{e}_R = (1,3) &\rightarrow & (1,\overline{3})_{R}(2,3)_{\phi}(2,\overline{3})_{\phi}(1,3)_L \nn \\
  \overline{E}_R = (2,1) &\rightarrow & (2,1)_{R}(2,\overline{3})_{\phi}(1,3)_ L
  \, ,
  \eea
and analogously for $E_L$ with other two possibilities
  \bea
\label{example2}
  \overline{e}^i_R = (1,1) &\rightarrow & (1,1)_ R(2,\overline{3})_{\phi}(2,3)_L \nn \\
  \overline{E}_R = (2,\overline{3}) &\rightarrow & (2,\overline{3})_{R}(2,3)_{\phi}(2,\overline{3})_{\phi}(2,3)_ L \, .
  \eea
In \eq{example2} we have indicated  the fields  by the indices: $\phi$ stands for the pseudo-Goldstone bosons, $R$ and $L$ for right-handed and left-handed leptons respectively. By comparing \eq{example1} with \eq{example2} we see that there is no choice for the  right-handed fermions representations that permits mixing between the first family charged lepton and the other two. As it happens,  we have the same problem in the neutrino sector, and  this means that it is
  impossible to reproduce the experimental lepton mixing matrix since we
  cannot have an angle different from zero between the first and
  the second family. In conclusion, we  are  forced to use only flavor singlets. Notice that we would arrive at the same conclusion
  if we had started  with a doublet composed by
  the first and the second family or if we had considered the
  quark sector.

 The previous  analysis shows that the introduction of a
 nonabelian flavor symmetry is not helpful since we are forced to use only
 flavor singlets as representations of the standard fermions if we want
 to reproduce the correct textures in the mixing matrices. Such a symmetry  could in principle be useful
  if  the  pseudo-Goldstone
  bosons content would be enlarged by making the flhiggs belong to different representations of $SU(3)$ and the flavor group, for instance, by having weak singlets in addition to doublets.
 Our aim is to build a model  as simple as possible and therefore we try avoiding such an enlargement. This leads us to  taking the abelian group $U(1)$ as our flavor symmetry.

\subsection{Spontaneous symmetry breaking}
\label{sec:ssb}

Our discussion so far has lead us to identify the low-energy symmetry we expect to see realized in the model  as $[SU(3)\times U(1)]_W\times U(1)_F$ plus the additional symmetry, which we take to be $U(1)_X$, that controls the exotic fermions.

 Once chosen the symmetry at the lower scale, we have to identify the minimal  global symmetry, the spontaneous breaking of which gives rise to the pseudo-Goldstone bosons to be identified with flavons and Higgs boson. Since we need at least two copies of $SU(3)\times U(1)$, plus two copies of an extra $U(1)$ to control the masses of the exotic fermions, we  end up with a group of rank 9, that we take to be
  $SU(10)$.

  The  $SU(10)$ global symmetry is spontaneously broken to
$SO(10)$ at the scale $f$.
This provides us with an effective theory with a cut off at the scale $\Lambda= 4\pi f$.
Fifty-four generators of  $SU(10)$ are
broken giving $54$ real Goldstone bosons we parametrize in a non-linear sigma model fashion as
\be
\Sigma (x) = \exp \left[ i \Pi (x) /f \right]
\Sigma_0 \, ,
\ee
with $\Pi(x)= t^a \pi^a(x)$, where  $t^a$ are
the broken generators of  $SU(10)$, $\pi^a(x)$ the fluctuations
around the vacuum $\Sigma_0$ given by
\be
\Sigma_0 \equiv \langle
\Sigma \rangle = \left( \begin{array}{cc|cc} 0 & I_{4\times 4}  & 0 & 0 \\ I_{4\times 4} & 0 & 0 & 0 \\
\hline
0 & 0 & 0 &1  \\ 0 & 0 &  1  & 0 \\
\end{array} \right) \, .
\label{vacuumsigma} 
\ee
The vacuum state (\ref{vacuumsigma}) can be rotated into its canonical form $I_{10\times10}$ by a change of basis. In this basis the breaking pattern is more evident but the sigma model dynamics more involved.

Within $SU(10)$ we identify seven subgroups
\be
SU(10) \supset  U(1)_F \times [SU(3) \times U(1)]_W^2 \times [U(1)_X]^2\, ,
 \ee
where the  $U(1)_F$ is the global flavor symmetry while the $[SU(3) \times U(1)]_W^2$
are  two copies of an extended electroweak gauge symmetry, the need of which we discussed in the previous section. The groups $[U(1)_X]^2$ are two copies of an extra gauge symmetry we need in order to separate standard fermions from the exotic fermions the model requires because of the enlarged $SU(3)$ symmetry that turns the weak doublets into triplets.

As discussed in sect.~\ref{sec:mb}, we want the flavor symmetry proper to be global so as not to have in the theory flavor charged gauge bosons that would make impossible for flavor and weak symmetry breaking to be of the same order. On the other hand, all the other symmetries in addition to those of the standard model are local so as to reduce the number of Goldstone bosons in the physical spectrum. 

The generators of the five $U(1)$ are taken to be
\bea
Y_{F_{1}} &=& \mbox{diag}\, (0,0,0,1,0,0,0,-1,0,0)/2 \nn \\
Y_{W_{1}} &=& \mbox{diag}\,(0,0,0,0,1,1,1,0,0,0)/\sqrt{6} \nn \\
Y_{W_{2}} &=& \mbox{diag}\,(1,1,1,0,0,0,0,0,0,0)/\sqrt{6} \,,
 \nn \\
Y_{X_{1}} &=& \mbox{diag}\,(0,0,0,0,0,0,0,0,1,0)/\sqrt{2} \nn \\
Y_{X_{2}} &=& \mbox{diag}\, (0,0,0,0,0,0,0,0,0,1)\sqrt{2} \,,
\eea
 while  the generators of the two copies
 of $SU(3)_{W}$ can be  identified with the corresponding generators
$Q^a_1$ and $Q^a_2$ with $a=1,...,8$  within $SU(10)$.
 Note that $Y_{F_{1}}$ and $Y_{W_{1,2}}$ are generators of $SU(10)$, while $Y_{X_{1,2}}$ are not and their normalization is chosen for simple convenience.

\begin{figure}\begin{center}\begin{picture}(260,100)(0,0)
\LongArrow(125,20)(140,20)
\LongArrow(80,80)(170,80)
\LongArrow(40,70)(40,30)
\LongArrow(220,70)(220,30)
\Text(40,20)[c]{$\left[SU(3)\,\times\, U(1)\right]^2_W\,\times\,U(1)_F\times\,[U(1)_X]^2$}\Text(220,20)[c]{$\left[SU(3)\,\times\, U(1)\right]_W\,\times\,U(1)_F\times\,U(1)_X$}
\Text(40,80)[c]{$SU(10)$}
\Text(220,80)[c]{$SO(10)$}
\end{picture}
\end{center}
\caption{Diagrammatic representation of the symmetry breaking structure of the sigma model.
Horizontal arrows indicate the spontaneous $SU(10)\to SO(10)$
global symmetry breaking, vertical arrows
the explicit breaking due to gauge interactions and plaquette terms (see the discussion in the text main body).}
\label{fig:symmetries}
\end{figure}

The breaking of $SU(10)$ into $SO(10)$ also breaks the subgroups
$[SU(3) \times U(1)]_W^2\times\,[U(1)_X]^2 $ and only a diagonal combination
survives. On the contrary, the flavor symmetry $U(1)_F$ survives the breaking and we eventually have that
 \be
  U(1)_F \times [SU(3)
\times U(1)]_W^2\times [U(1)_X]^2 \rightarrow U(1)_F \times [SU(3) \times U(1)]_W \times U(1)_X
\, .
\ee
The breaking in the gauge sector $[SU(3) \times U(1)]_W^2 \times [U(1)_X]^2
\rightarrow[SU(3) \times  U(1)]_W \times U(1)_X $ leaves $10$ gauge bosons
massive after eating $10$ of the $54$ real Goldstone bosons. The remaining
$44$ Goldstone bosons can be labeled according to  representations of the $U(1)_F \times
[SU(3) \times U(1)]_W \times U(1)_X $ symmetry:
\begin{itemize}
\item 2 complex fields $\Phi_1 \,[3_{(1,1/2,0)}]$ and
$\Phi_2 \,[3_{(1,-1/2,0)}]$, accounting for 12 degrees of freedom. They transform as  triplets of $[SU(3)]_W$ and have the same  $U(1)_W$
and opposite $U(1)_F$ charges. They are not charged under the exotic gauge symmetry  $U(1)_X$.
\item 2 complex fields $\Phi_3 \,[3_{(1,0,1/2)}]$ and
$\Phi_4 \,[3_{(1,0,-1/2)}]$, accounting for other 12 degrees of freedom.  They
transform as  triplets of $[SU(3)]_W$ and have the same  $U(1)_W$
and opposite $U(1)_X$ charge. They are not charged under the flavor symmetry  $U(1)_F$.
 \item a sextet of complex fields $z_{ij} \, [6_{(2,0,0)}]$,  for 12 degrees of freedom,
\item 4 complex fields $s\, [1_{(0,-1,0)}]$, $s_1\, [1_{(0,-1/2,1/2)}]$, $s_2\, [1_{(0,1/2,1/2)}]$, $s_3\, [1_{(0,0,-1)}]$, for for the remaining 8 degrees of freedom.
\end{itemize}
In the above notation, the
representations with respect to the $SU(3)_W$ are indicated
between square brackets and the indexes are the $U(1)$ charges: the
first  refers to the weak group, the second to the flavor
group and the third to the exotic.

In terms of these representations, the field  $\Pi(x)$ can then be written as
\be
\Pi (x) = \left( \begin{array}{ccc|c|ccc|c|cc}
0& 0& 0 &  & &  &  & & & \\
0& 0& 0 & \Phi_1/\sqrt{2} & & z_{ij} & &\Phi_2/\sqrt{2} &\Phi_3/\sqrt{2}&\Phi_4/\sqrt{2}\\
0& 0& 0 &  & &  &  & & & \\
\hline
&\Phi_1^*/\sqrt{2}  &  & 0 & &\Phi_2/\sqrt{2}  &  & s & s_1/2 & s_2/2\\
\hline
 & &  &  &0 & 0 & 0 & & & \\
& z_{ij}^*&  &  \Phi_2^*/\sqrt{2}& 0&  0&0  &  \Phi_1^*/\sqrt{2}&\Phi_4^*/\sqrt{2}& \Phi_3^*/\sqrt{2}\\
  & &  &  &0 & 0 & 0 & & & \\
\hline
 &\Phi_2^*/\sqrt{2}  &  & s^* & &\Phi_1/\sqrt{2}  &  & 0 & s_2^*/2 & s_1^*/2\\
 \hline
&\Phi_3^* /\sqrt{2} &  & s_1^*/2 & &\Phi_4/\sqrt{2}  &  & s_2/2 & 0 & s_3\\
 &\Phi_4^* /\sqrt{2} &  & s_2^*/2 & &\Phi_3/\sqrt{2}  &  & s_1/2 & s_3^*&0\\
 \end{array} \right) \, ,
\label{pione}
\ee
where we have put zeros for the components that are going to be eaten by the gauge fields becoming massive.
All these fields are still Goldstone bosons with no potential; their potential arises after the explicit breaking of the symmetry to which we now turn.

\subsection{Explicit collective symmetry breaking}
\label{sec:esb}

 The effective lagrangian of the pseudo-Goldstone
bosons must contain terms that explicitly  break the $SU(10)$ global
symmetry. These terms provide masses of the order of the scale $f$ for the $s,\, s_i$ and $z_{ij}$ fields. However, each term separately preserve  enough symmetry to keep
the flhiggs fields  $\Phi_i$  exact Goldstone  bosons.
Only the simultaneous action of two or more of the terms (collective breaking)
 turns them into pseudo-Goldstone with  a potential, even though there is still no mass term. Quadratic terms for the flhiggs will come from the coupling to right-handed neutrino, as we shall discuss presently.

The effective lagrangian is given by the kinetic term
\be
\mathcal{L}_0 = \frac{f^2}{2}\Tr (D^\mu \Sigma) (D_\mu \Sigma)^* \, ,
\ee
the covariant derivative of which couples the pseudo-Goldstone bosons to the gauge fields:
\be
D_\mu \Sigma =
\partial_\mu + i g_i W^a _{i_{\mu}}(Q^a_i \Sigma + \Sigma
Q_i^{a,T}) + i g'_i y_i B_{i_{\mu}}(Y_{W_{i}} \Sigma + \Sigma Y_{W_{i}}^{T})
+ i k_i X_{i_{\mu}}(Y_{X_{i}} \Sigma + \Sigma Y_{X_{i}}^{T}) \quad i=1,2 \,, \label{kinetic}
\ee
where $W^a _{i_{\mu}}$, $B_{i_{\mu}}$ and $X_{i_{\mu}}$ are
the gauge bosons of the $SU(3)_{_W{i}}$ , $U(1)_{_W{i}}$ and $U(1)_{_X{i}}$
respectively , $Q^a_i$, $Y_{W_{i}}$ and $Y_{X_{i}}$ their generators and $y_i$ the  $U(1)_{_W{i}}$ charges.

The lagrangian in \eq{kinetic}  gives mass to the $z_{ij}$ and  $s_3$  fields. On the other hand,
each  term of index $i$ preserves a $SU(3)$ symmetry so that  only when taken together they can give a contribution to the potential of the flhiggs fields.

At this point the fields $s$, $s_1$ and $s_2$ are still massless. They play no important role in the model but cannot remain massless. To give them a mass, we introduce
plaquette terms---terms made out of components of the $\Sigma$ field that preserve enough symmetry not to induce masses for the flhiggs fields. 

As an example, one of these plaquette term can be written by looking at the Goldstone fields in the matrix \eq{pione} after having rotate it by the vacuum $\Sigma_0$.  We select the field $s^*$ to which we want to give mass in the components $(8,\, 8)$ and  $(4,\, 4)$ of the matrix $\Sigma_0 \Pi (x)$. Both these choices leave a different  $SU(9)$ symmetry acting on the remaining columns and rows that then prevents further terms to the potential of the fields that transform in the coset of $SU(10)/SU(9)$.

The other possible plaquette terms are given by choosing by the same token the two couples  $(4,10)$ and  $(8,9)$ and   $(4,9)$ and  $(8,10)$  components to give masses to the $s_1$ and $s_2$ respectively. Together they  induce  (harmless) terms and corrections into the coefficients of the flhiggs potential.

After adding the plaquette terms, we therefore have  the effective lagrangian
 \bea
\mathcal{L}& =& \mathcal{L}_0 + a_1^2 f^2 \Sigma_{4,4}\Sigma^*_{4,4}
+ a_2^2 f^2\Sigma_{8,8}\Sigma^*_{8,8}
 + a_3^2 f^2 \Sigma_{4,9}\Sigma^*_{4,9} \nn \\
&+&  a_4^2 f^2 \Sigma_{8,10}\Sigma^*_{8,10}
+ a_5^2 f^2 \Sigma_{4,10}\Sigma^*_{4,10}
+ a_6^2 f^2 \Sigma_{8,9}\Sigma^*_{8,9}\,, \label{efl}
 \eea
where $a_{i}$ are  coefficients of $\mathcal{O}(1)$. The relative signs of the plaquette terms are in principle arbitrary and presumably fixed by the ultraviolet  completion of the theory. At this level we simply require $m^2_{s_{i}}>0$.

In \sec{sec:ssb} we said that in the breaking of $[SU(3) \times
U(1)]_W^2 \rightarrow [SU(3) \times U(1)]_W $ nine gauge bosons
become massive. We now see that their masses are given by
\be
M^2_{W'_a}=\frac{(g_1^2 + g_2^2)}{2}f^2 \quad M^2_{B'}= \frac{({g'}_1^2 +
{g'}_2^2)}{2}f^2  \quad M^2_{X'}= \frac{({k}_1^2 + {k}_2^2)}{2} f^2 \,,
\label{heavy}
 \ee
where $a=1, \dots ,8$.

These  heavy gauge bosons---because of their mixing with those with lighter masses to be identified with the standard model gauge bosons---induce corrections on many observables that we know to be constrained by high-precision measurements, mainly coming from low-energy physics (like atomic parity violation and neutrino-hadron scattering). Their presence is the major constrain on the scale $f$ and, accordingly, the naturalness of the model, as discussed for the littlest-Higgs model in \cite{littlehiggs2}. We shall come back to them when we discuss these constrains in the flhiggs model in section \ref{sec:gbc}.

 The
effective potential for the flhiggs fields is given by the tree-level contribution coming from the
plaquettes and the one-loop Coleman-Weinberg effective potential arising from the gauge interactions:
 \be
 \label{cw}
\frac{\Lambda^2}{16\pi ^2}\Tr [M^2(\Sigma)] + \frac{3}{64 \pi^2}\Tr \left[M^4(\Sigma)\left(\log\frac{M^2(\Sigma)}{\Lambda^2} + {\rm const.} \right)\right] \, ,
 \ee
 where the second, logarithmic terms is very much suppressed and is not included in what follows.

The effective
potential ${ O}(f^{-2})$  is obtained by expanding the sigma-model field $\Sigma$ and is given by
\bea
\label{pot}
\mathcal{V}_0 [\Phi_i, z_{ij}, s, s_i] &=&
\frac{2}{3}\,g_1^2\, f^2 |\frac{z_{ij}}{\sqrt{2}} -\frac{i}{2\sqrt{2} f} \big(\Phi_{1_{i}}
\Phi_{2_{j}}+\Phi_{3_{i}}
\Phi_{4_{j}}\big) |^2 +\frac{2}{3}\, g_2^2\, f^2 |\frac{z_{ij}}{\sqrt{2}}
+\frac{i}{2\sqrt{2}  f} (\Phi_{1_{i}} \Phi_{2_{j}}+\Phi_{3_{i}}
\Phi_{4_{j}}\big)|^2 \nn \\
&+& \frac{1}{3}\,{g'}_1^2\, f^2 |\frac{z_{ij}}{\sqrt{2}} -\frac{i}{2\sqrt{2} f} \big(\Phi_{1_{i}}
\Phi_{2_{j}}+\Phi_{3_{i}}\Phi_{4_{j}} \big)|^2 +\frac{1}{3}\, {g'}_2^2\, f^2 |\frac{z_{ij}}{\sqrt{2}}
+\frac{i}{2\sqrt{2} f} (\Phi_{1_{i}} \Phi_{2_{j}}+\Phi_{3_{i}}
\Phi_{4_{j}}\big)|^2 \nn \\
&+& \frac{1}{2}\,k_1^2\, f^2 |s_3 -\frac{i}{2 f} \big(\frac{s_2 s_1^*}{2} +
\Phi_{3}^{\dag} \Phi_{4}\big) |^2 +\frac{1}{2}\,
k_2^2\, f^2 |s_3 +\frac{i}{2 f} \big(\frac{s_2 s_1^*}{2} +
\Phi_{3}^{\dag}
\Phi_{4}\big) |^2  \\
&+& a_1^2\, f^2\, |s -\frac{i}{2 f} \big(\frac{ s_1 s_2}{2}
+\Phi^{\dag}_{1} \Phi_{2}\big)|^2 + a_2^2\, f^2\, |s + \frac{i}{2
f} \big( \frac{s_1 s_2}{2} +\Phi^{\dag}_{1} \Phi_{2}\big)|^2 \nn\\
&+& \frac{1}{4}\,a_3^2\, f^2\, |s_1 -\frac{i}{2 f} \big( s s^*_2+ s_2
s_3^* +\Phi^{\dag}_{1} \Phi_{3}+ \Phi^{\dag}_{4}
\Phi_{2}\big)|^2
+ \frac{1}{4}\, a_4^2\, f^2\, |s_1 + \frac{i}{2 f}
\big(s s^*_2+ s_2 s_3^* +\Phi^{\dag}_{1} \Phi_{3}+
\Phi^{\dag}_{4} \Phi_{2} \big)|^2
\nn\\
&+& \frac{1}{4}\, a_5^2\, f^2\, |s_2 -\frac{i}{2 f} \big( s s^*_1+ s_1
s_3 +\Phi^{\dag}_{1} \Phi_{4}+ \Phi^{\dag}_{3}
\Phi_{2}\big)|^2
+ \frac{1}{4}\, a_6^2\, f^2\, |s_2 + \frac{i}{2 f} \big(s
s^*_1+ s_1 s_3 +\Phi^{\dag}_{1} \Phi_{4}+
\Phi^{\dag}_{3} \Phi_{2} \big)|^2 \,. \nn
\eea

 From \eq{pot} we see by inspection that the  effective potential  gives mass to the scalar fields
 $s$, $s_1$, $s_2$, $s_3$ and $z$, their masses  given by
\bea
 m^2_z & =& \frac{( 2 g_1^2+ 2 g_2^2+ {g'}_1^2 + {g'}_2^2 )}{6} f^2
 \quad \quad  m^2_{s_{3}} =\frac{ (k_1^2 + k_2^2)}{2} f^2 \nn \\
m^2_s & =& (a_1^2 + a_2^2) f^2  \quad \quad m^2_{s_{1}}=
\frac{(a_3^2 + a_4^2)}{2} f^2  \quad \quad m^2_{s_{2}}= \frac{(a_5^2 + a_6^2)}{2} f^2
\,, \eea
respectively. The effect of these states must be included in the study of the low-energy observables together with that of the heavy gauge bosons.  

After integrating out the massive states by means of their equations of motion,
the potential of the four pseudo-Goldstone bosons $\Phi_{i}$, the flhiggs, is made of only quartic terms
 \bea
\mathcal{V}_1 [\Phi_i ] &=&\lambda_1 (\Phi_1^{\dag}\Phi_1)(\Phi_2^{\dag}\Phi_2) +\lambda_2 (\Phi_3^{\dag}\Phi_3)(\Phi_4^{\dag}\Phi_4)
+ \lambda_3
(\Phi_1^{\dag}\Phi_1)(\Phi_3^{\dag}\Phi_3)+\lambda_4
(\Phi_2^{\dag}\Phi_2)(\Phi_4^{\dag}\Phi_4) \nn\\
&+& \lambda_5
(\Phi_1^{\dag}\Phi_1)(\Phi_4^{\dag}\Phi_4)+\lambda_6
(\Phi_2^{\dag}\Phi_2)(\Phi_3^{\dag}\Phi_3) \\
 &+&   \xi_1 |\Phi_1^{\dag}\Phi_2|^2+ \xi_2 |\Phi_3^{\dag}\Phi_4|^2
+ \xi_3(\Phi_1^{\dag}\Phi_3)(\Phi_2^{\dag}\Phi_4) + \xi_4  (\Phi_1^{\dag} \Phi_4)(\Phi_2^{\dag}\Phi_3)\,,  \nn  \label{pot1}
 \eea
 where the coefficients are given by
 \be
 \lambda_1 =\lambda_2= \frac{(2 g_1^2+{g'}_1^{2}) (2 g_2^2+{g'}_2^{2})}{2 g_1^2+ {g'}_1^{2}
+2  g_2^2+ {g'}_2^{2}} \quad \quad \lambda_3 = \lambda_4=\frac{
a_3^2 a_4^{2}}{ a_3^2+ a_4^{2}}
\quad \quad \lambda_5 = \lambda_6 =\frac{a_5^2 a_6^{2}}{
a_5^2+ a_6^{2}}
\ee
and
\bea
 \xi_1  &=& \frac{(2
g_1^2+{g'}_1^{2}) (2 g_2^2+{g'}_2^{2})}{2 g_1^2+ {g'}_1^{2} + 2
g_2^2+ {g'}_2^{2}} + \frac{a_1^2 a_2^2}{a_1^2+a_2^2}
\quad \quad
\xi_2 =\frac{k_1^2 k_2^2}{k_1^2+k_2^2} + \frac{a_5^2 a_6^{2}}{a_5^2+ a_6^{2}} \\
 \xi_3  &=& \frac{(2
g_1^2+{g'}_1^{2}) (2 g_2^2+{g'}_2^{2})}{2 g_1^2+ {g'}_1^{2} + 2
g_2^2+ {g'}_2^{2}} + \frac{a_3^2 a_4^2}{a_3^2+a_4^2}
\quad \quad
\xi_4 = \frac{(2
g_1^2+{g'}_1^{2}) (2 g_2^2+{g'}_2^{2})}{2 g_1^2+ {g'}_1^{2} + 2
g_2^2+ {g'}_2^{2}} + \frac{a_5^2 a_6^{2}}{a_5^2+ a_6^{2}}  \, .
 \eea
The coefficients $\xi_1$, $\xi_3$ and $\xi_4$ differ only by the plaquette contributions. Notice that we can take them equal if we assume the plaquette coefficients to be equal as well.

 Quadratic terms
   \be
\mathcal{V}_2 [\Phi_i ] =  \mu^2_1 (\Phi_1^{\dag}\Phi_1) + \mu^2_2 (\Phi_2^{\dag}\Phi_2)+\mu^2_3 (\Phi_3^{\dag}\Phi_3)
   + \mu^2_4 (\Phi_4^{\dag}\Phi_4) \label{pot2}
   \ee
    that are necessary to induce vacuum expectation values for the flhiggs fields,
   and quartic terms of the type
  \be
 \mathcal{V}_3 [\Phi_i ] =  \chi_1 (\Phi_1^{\dag}\Phi_1)^2 +\chi_2(\Phi_2^{\dag}\Phi_2)^2 +\chi_3
  (\Phi_3^{\dag}\Phi_3)^2 +\chi_4 (\Phi_4^{\dag}\Phi_4)^2 \label{pot3}
 \ee
are not generated at one--loop in the bosonic sector discussed so far.
 In order to introduce them we  couple the pseudo-Goldstone bosons to right-handed neutrinos with masses at the scale $f$. This means that the flavor and electroweak symmetry breaking of the model is triggered by the right-handed neutrinos. This is done again along the lines of the little-Higgs collective symmetry breaking:
 to prevent  quadratically
 divergent mass term for $\Phi_i$---and thus render useless what done up to this point---the Yukawa lagrangian of the
 right-handed neutrinos sector is constructed by terms that taken separately leave invariant some subgroups of the
 approximate global symmetry $SU(10)$. In this way the  flhiggs bosons receive a mass
 term only from diagrams in which all the approximate global
 symmetries of the Yukawa lagrangian are broken. Because of this collective breaking,  the one-loop
 contributions to the flhiggs masses are only logarithmic
 divergent.

 The right handed neutrino sector is given by sixteen 10-components multiplets
\bea
\label{neuright}
\begin{array}{cccc}
 N^1_R =\left( \begin{array}{c} 0_\alpha \\ \nu^1_R \\0_\alpha \\0\\0\\0
        \end{array}\right)
 & N^2_R =\left( \begin{array}{c} 0_\alpha \\ 0\\0\\\nu_R^{2} \\0 \\0
        \end{array}\right)
 &N^3_R =\left( \begin{array}{c} 0_\alpha \\ 0\\0\\0_\alpha\\\nu_R^{3} \\0
        \end{array}\right)
 &N^4_R =\left( \begin{array}{c} 0_\alpha \\ 0\\0\\0\\0_\alpha\\\nu_R^{4}
        \end{array}\right)
        \end{array}& \nn\\
 &&\nn\\
\begin{array}{cccc}
 N^5_R =\left( \begin{array}{c}
\nu_{R,\alpha}^{5}\\0
\\{\nu'}_{R,\alpha}^{5} \\\tilde{\nu'}_{R}^{5}\\\hat{\nu}_{R}^{5}\\\hat{\nu'}_{R}^{5}
        \end{array}\right)
& N^6_R= \left(\begin{array}{c}
\nu_{R,\alpha}^{6}\\\tilde{\nu}_{R}^{6}
\\{\nu'}_{R,\alpha}^{6} \\0\\\hat{\nu}_{R}^{6}\\\hat{\nu'}_{R}^{6}
        \end{array}\right)
&N^7_R = \left(\begin{array}{c}
\nu_{R,\alpha}^{7}\\\tilde{\nu}_{R}^{7}
\\{\nu'}_{R,\alpha}^{7} \\\tilde{\nu'}_{R}^{7}\\0\\\hat{\nu'}_{R}^{7}
        \end{array}\right)
 &N^8_R = \left(\begin{array}{c} \nu_{R,\alpha}^{8}\\\tilde{\nu}_{R}^{8}
\\{\nu'}_{R,\alpha}^{8}
\\\tilde{\nu'}_{R}^{8}\\\hat{\nu}_{R}^{8}\\0
        \end{array}\right)
 \end{array}& \nn\\
 &&\nn\\
\begin{array}{cccc}
 N^9_R =\left( \begin{array}{c}
\nu_{R,\alpha}^{9}\\0
\\0 \\0\\\hat{\nu}_{R}^{9}\\0
        \end{array}\right)
&N^{10}_R =\left( \begin{array}{c} \nu_{R,\alpha}^{10}\\0
\\0 \\0\\\hat{\nu}_{R}^{10}\\0
        \end{array}\right)
&N^{11}_R =\left( \begin{array}{c} 0_\alpha\\0\\{\nu'}_{R,\alpha}^{11}\\0
\\0 \\\hat{\nu'}_{R}^{11}
        \end{array}\right)
&N^{12}_R =\left( \begin{array}{c} 0_\alpha\\0\\{\nu'}_{R,\alpha}^{12}\\0
\\0 \\\hat{\nu'}_{R}^{12}
        \end{array}\right)
\end{array}&\nn\\
 &&\nn\\
\begin{array}{cccc}
 N^{13}_R =\left( \begin{array}{c}
\nu_{R,\alpha}^{13}\\ \tilde{\nu}_{R}^{13}\\0\\0\\0\\0
        \end{array}\right)
&N^{14}_R =\left( \begin{array}{c} \nu_{R,\alpha}^{14}\\\tilde{\nu}_{R}^{14}\\0\\0\\0\\0
        \end{array}\right)
&N^{15}_R =\left( \begin{array}{c} 0_\alpha\\0\\{\nu'}_{R,\alpha}^{15} \\\tilde{\nu'}_{R}^{15}\\0\\0
        \end{array}\right)
&N^{16}_R =\left( \begin{array}{c} 0_\alpha\\0\\{\nu'}_{R,\alpha}^{16}\\\tilde{\nu'}_{R}^{15}\\0\\0
        \end{array}\right)
\end{array}
 \,,
\eea
where $\alpha=1,2,3$ and  $0_\alpha = (0,0,0)^T$.

Only the $N^i_{R}$ with $a=1,\cdots,8$ couple directly to the fermions.
The reason why we introduce so many fields is that we eventually want  different, and independent, mass terms $\mu_i$  to be induced in the effective potential by the right-handed neutrino sector and also we do not want right-handed neutrinos massless.

 The Yukawa lagrangian for the right-handed neutrinos can be written in a $SU(10)$-invariant manner as
 \bea \label{yukright1}
\mathcal{L}_Y^{\nu_{R}}&=& \eta_1 f \big( \overline{N^{1^{c}}_L}\Sigma  N^5_R\big) +
\eta_2 f \big( \overline{N^{2^{c}}_L}\Sigma  N^6_R\big) + \eta_3 f \big(
\overline{N^{3^{c}}_L}\Sigma  N^7_R\big) + \eta_4 f \big(\overline{ N^{4^{c}}_L}\Sigma
N^8_R\big)\nn\\
&+& \eta_5 f \big( \overline{N^{9^{c}}_L}\,  N^5_R\big)+ \eta_6 f \big(
\overline{N^{11^{c}}_L}\,  N^5_R\big)+
\eta_7 f \big(\overline{ N^{10^{c}}_L}\,  N^6_R\big)+ \eta_8 f \big(\overline{ N^{12^{c}}_L}\,  N^6_R\big)\nn \\
&+& \eta_9 f \big(\overline{ N^{13^{c}}_L}\,  N^7_R\big)+ \eta_{10} f \big(
\overline{N^{15^{c}}_L}\,  N^7_R\big)+ \eta_{11} f \big(\overline{ N^{14^{c}}_L}\,
N^8_R\big)+ \eta_{12} f \big( \overline{N^{16^{c}}_L}\,  N^8_R\big) \,.
\eea
In \eq{yukright1} all the terms leave invariant  different subgroups
of $SU(10)$: the first four, four different $SU(9)$ global
symmetries---easily identifiable by the zeros in $N_{1-4}$---the remaining  eight different $SU(6)$---to be identified by the zeros in $N_{9-12}$.

Substituting in \eq{yukright1} the expression ${ O}(f^{-2})$ for $\Sigma$---as
given in \eq{pione}---and for the right-handed neutrino multiplets
the expression given in \eq{neuright}, we obtain the leading order lagrangian
 \bea \label{yukright2}
\mathcal{L}_Y^{\nu_{R}}&=&
\eta_1 f \Big[\overline{\nu_L^{1c}} \tilde{\nu'}_R^5
\Big(1-\frac{\Phi_1^{\dag}\Phi_1}{2f^2}-
\frac{\Phi_2^{\dag}\Phi_2}{2f^2}- \frac{(2 s s^* +s_1 s_1^* + s_2 s_2^*)}{2f^2}\Big) +  \frac{i}{f}\overline{\nu_L^{1c}} \Big(\Phi_2^T \nu_R^5  + \Phi_1^{\dag} {\nu'}_R^5+
s_1\hat{\nu'}_{R}^{5}+ s_2\hat{\nu}_{R}^{5}  \Big) \Big] \nn \\
&+& \eta_2 f \Big[ \overline{\nu_L^{2c}} \tilde{\nu}_R^6
\Big(1-\frac{\Phi_1^{\dag}\Phi_1}{2f^2}-
\frac{\Phi_2^{\dag}\Phi_2}{2f^2}- \frac{(2 s s^* +s_1 s_1^* + s_2 s_2^*)}{2f^2}\Big)+ \frac{i}{f}\overline{\nu_L^{2c}} \Big(\Phi_1^T \nu_R^6  + \Phi_2^{\dag}
{\nu'}_R^6+
s_2^*\hat{\nu'}_{R}^{6}+ s_1^*\hat{\nu}_{R}^{6}  \Big)\Big] \nn \\
&+& \eta_3 f \Big[ \overline{\nu_L^{3c}} \hat{\nu'}_R^7
\Big(1-\frac{\Phi_3^{\dag}\Phi_3}{2f^2}-
\frac{\Phi_4^{\dag}\Phi_4}{2f^2}- \frac{( 2 s_3 s_3^* +s_1 s_1^* + s_2 s_2^*)}{2f^2}\Big) + \frac{i}{f}\overline{\nu_L^{3c}} \Big(\Phi_4^T \nu_R^7  +
\Phi_3^{\dag} {\nu'}_R^7+
s_1^*\tilde{\nu'}_{R}^{7}+ s_2\tilde{\nu}_{R}^{7}  \Big)\Big] \nn \\
&+& \eta_4 f \Big[\overline{ \nu_L^{4c}} \hat{\nu}_R^8
\Big(1-\frac{\Phi_3^{\dag}\Phi_3}{2f^2}-
\frac{\Phi_4^{\dag}\Phi_4}{2f^2}- \frac{( 2 s_3 s_3^* +s_1 s_1^* + s_2 s_2^*)}{2f^2}\Big)+ \frac{i}{f}\overline{\nu_L^{4c}} \Big(\Phi_3^T \nu_R^8  +
\Phi_4^{\dag} {\nu'}_R^8+
s_2^*\tilde{\nu'}_{R}^{8}+ s_1\tilde{\nu}_{R}^{8}  \Big)\Big]+ \nn \\
&+& \eta_5 f \Big[\overline{\hat{\nu}_L^{9c}} \hat{\nu}_R^5+
\overline{\nu_L^{9c}} \nu_R^5\Big] + \eta_6 f \Big[\overline{\hat{\nu'}_L^{11c}}
\hat{\nu'}_R^5+\overline{{\nu'}_L^{11c}} {\nu'}_R^5\Big] + \eta_7 f
\Big[\overline{\hat{\nu}_L^{10c}} \hat{\nu}_R^6+ \overline{\nu_L^{10c}}
\nu_R^6\Big] + \eta_8 f \Big[\overline{\hat{\nu'}_L^{10c}}
\hat{\nu'}_R^6+\overline{\nu_L^{10c}} {\nu'}_R^6\Big] \nn\\
&+& \eta_9 f \Big[\overline{\tilde{\nu}_L^{13c}} \tilde{\nu}_R^7+\overline{\nu_L^{13c}} \nu_R^7 \Big] + \eta_{10} f
\Big[\overline{\tilde{\nu'}_L^{15c}}
\tilde{\nu'}_R^7+\overline{{\nu'}_L^{15c}} {\nu'}_R^7   \Big]+\eta_{11} f \Big[\overline{\tilde{\nu}_L^{14c}} \tilde{\nu}_R^8+\overline{\nu_L^{14c}} \nu_R^8\Big]+ \eta_{12} f
\Big[ \overline{\tilde{\nu'}_L^{16c}}
\tilde{\nu'}_R^8+\overline{{\nu'}_L^{16c}} {\nu'}_R^8\Big]\,.
 \eea
 From \eq{yukright2}
we see that, after integrating out the neutrinos, the divergent one-loop contributions to the
pseudo-Goldstone bosons masses in the effective potential ${\cal V}_2[\Phi_i]$ of \eq{pot2} are given by
\bea
 \label{massephi}
\mu^2_1 &\simeq& (\eta_1^2\eta^2_5 +\eta_2^2\eta^2_7 ) 
\frac{ f^2}{(4\pi)^2} \log
\frac{\Lambda^2}{M_\eta^2}  \nn \\
 \mu^2_2 &\simeq&
(\eta_1^2\eta^2_6 +\eta_2^2\eta^2_7 ) 
\frac{ f^2}{(4\pi)^2}
\log \frac{\Lambda^2}{M_\eta^2} \nn\\
 \mu^2_3 &\simeq& (\eta_3^2\eta^2_{10} +\eta_4^2\eta^2_{11} )
\frac{ f^2}{(4\pi)^2} \log
\frac{\Lambda^2}{M_\eta^2}  \nn \\
 \mu^2_4 &\simeq&
(\eta_3^2\eta^2_9 +\eta_4^2\eta^2_{12} )
\frac{ f^2}{(4\pi)^2} \log \frac{\Lambda^2}{M_\eta^2} \,,
\eea where in the logarithm of \eq{massephi} we have generically
indicated the mass of right handed neutrinos with $M_\eta \simeq
\eta f$. The one-loop quadratically divergent contributions are cancelled by the collective symmetry breaking and the masses are always proportional to two of the coefficients $\eta$.

From \eq{yukright2}, we can also estimate the one-loop
divergent contributions to the quartic terms  in the effective potential ${\cal V}_3[\Phi_i]$ of\eq{pot3} coming
 from the right-handed neutrino sector. The coefficients $\chi_i$ turn out to be logarithmically divergent and proportional to four, not necessarily different, powers of $\eta$:
 \be
 \chi_{i} \simeq \eta_{k,j}^4 \frac{f^2}{(4\pi)^2} \log
\frac{\Lambda^2}{M_\eta^2}  \,.
 \ee
They only play a minor role in what follows.


\subsection{Vacuum expectation value}
\label{sec:vev}

The effective potential for the pseudo Goldstone bosons is therefore made of the sum of eqs.\ (\ref{pot1}),
(\ref{pot2}) and (\ref{pot3})
\be \mathcal{V}[\Phi_i] =
\mathcal{V}_1+\mathcal{V}_2+\mathcal{V}_3 \,. \label{pot-final}
\ee
We want to find  vacuum expectation values for the flhiggs fields $\Phi_i$ in this potential that breaks the symmetry $[SU(3)\times
 U(1)]_W\times U(1)_F\times U(1)_X$ down to the electric charge group $U(1)_Q$.  Such a  vacuum is, in general,  given  by the field configurations
 \be
\vev{\Phi_1}= \left(\begin{array}{c}
  0 \\
 v_W /2\\
  v_{F_{1}}/2 \\
\end{array}\right)
\quad \quad
\vev{\Phi_2}= \left(\begin{array}{c}
  0 \\
 v_W /2\\
  v_{F_{2}}/2 \\
\end{array}\right)
\quad \quad
 \vev{\Phi_3}= \left(\begin{array}{c}
  0 \\
  0 \\
  v_{X_{1}}/2\\
\end{array}\right)
\quad \quad
\vev{\Phi_4}= \left(\begin{array}{c}
  0 \\
  0 \\
 v_{X_{2}}/2 \\
\end{array}\right)\,. \label{vacuum}
 \ee

The conditions to be satisfied, in order for \eq{vacuum} to be a  minimum, are the vanishing of the $24$ first derivatives:
\be
\left. \frac{\partial V[\Phi]}{\partial \Phi_i}\right|_{\Phi_i=\langle \Phi_i\rangle} \,. \label{der}
\ee
 Substituting the field
configuration of \eq{vacuum} in \eq{der}, we
have  $16$ equations satisfied and eight  conditions that
$v_W,\,v_{F1},\,v_{F2},\,v_{X1},\,v_{X2}$ and the parameters of the
potential must satisfy. Among them we have the following two
equations
\bea \xi_4 v_{F1}+ \xi_3
v_{F2}&=&0 \nn\\
\xi_3 v_{F1}+ \xi_4 v_{F2}&=&0\,.
\eea
We take the
solution in which $\xi_4= \xi_3$ and $v_{F2}=-v_{F1}$.  This solution is quite natural if, as pointed out in sec. \ref{sec:esb}, we consider all plaquette terms to come  with equal strengths. We also impose for simplicity that
$v_{X1}= v_{X2}=v_X$ and $v_{F1}=v_{F2}=v_F$. The values of $v_{X}$
and $v_F$ need not be equal but we shall identify them to obtain a
  model with only two vacuum values and simpler expressions for them in terms of the parameters.
  On the other hand, we do want to keep $v_F$ distinct from $v_W$
  because otherwise the increase in symmetry would lead to the presence of extra Goldstone bosons and other undesirable   phenomenological consequences for the model.
  
Under these assumptions, the field configuration of \eq{vacuum} becomes
\be
\vev{\Phi_1}= \left(\begin{array}{c}
  0 \\
 v_W /2\\
  v_{F}/2 \\
\end{array}\right)
\quad \quad \vev{\Phi_2}= \left(\begin{array}{c}
  0 \\
 v_W /2\\
 - v_{F}/2 \\
\end{array}\right)
\quad \quad
 \vev{\Phi_3}= \left(\begin{array}{c}
  0 \\
  0 \\
  v_{F}/2\\
\end{array}\right)
\quad \quad \vev{\Phi_4}= \left(\begin{array}{c}
  0 \\
  0 \\
 v_{F}/2 \\
\end{array}\right)\,, \label{vacuumdue}
 \ee
that is the vacuum expectation value we are going to use in what follows.

At this point we are left with six independent conditions that reduce to four if 
\be
\label{xi3}
\xi_3 = \Big(1-\frac{v_W^2}{v_F^2} \Big) \xi_1 \,,
\ee

The four remaining equations  yield the following expressions for the vacua
as function of the coefficients of the effective potential:
\bea
\label{vevwf}
v_W^2 & = &  \frac{(\lambda_3 +\lambda_6 - \lambda_4 -\lambda_5) (\mu_1^2 - \mu_2^2 + \mu_3^2 -\mu_4^2)  +2 (\chi_2 -\chi_1)(\mu_4^2-\mu_3^2) + 2 (\chi_4 -\chi_3)(\mu_1^2-\mu_2^2)}{ 4 (\chi_1-\chi_2)(\chi_3-\chi_4) - (\lambda_4-\lambda_3)^2 -(\lambda_5-\lambda_6)^2)}  \nn \\
v_F^2 & = & \frac{(\lambda_3 +\lambda_6 - \lambda_4 -\lambda_5)
(\mu_1^2 - \mu_2^2) + 2 (\chi_2 -\chi_1) (\mu_3^2-\mu_4^2)}{4
(\chi_1-\chi_2)(\chi_3-\chi_4) - (\lambda_4-\lambda_3)^2
-(\lambda_5-\lambda_6)^2} \,,
\eea
and also
yield  the following conditions on the $\mu_i^2$ since we have reduced
the number of degrees of freedom by imposing the previous equalities.

\bea
\label{condmui}
& -&\frac{-\mu_2^2(2 \xi_1-\xi_2  - 2 \chi_3-\lambda_2-\lambda_6-\lambda_3 )+\mu_3^2(\xi_1- 2 \chi_2 -\lambda_1-\lambda_4-\lambda_6  )}{(\lambda_1+\xi_1 + 2\chi_2)(2 \xi_1-\xi_2  - 2 \chi_3-\lambda_2-\lambda_6-\lambda_3)-(\lambda_6+ \lambda_3)(\xi_1- 2 \chi_2 -\lambda_1-\lambda_4-\lambda_6 )}
  \nn \\
&+&\frac{(-\mu_1^2+\mu_2^2)(-2 \chi_3 + 2 \chi_4 +\lambda_4 +\lambda_5 -\lambda_6-\lambda_3) + (\mu_3^2-\mu_4^2)(-2 \chi_1+2\chi_2+\lambda_4-\lambda_5+ \lambda_6-\lambda_3 )}{(\lambda_4+\lambda_5-\lambda_6 -\lambda_3)(-2 \chi_1+2\chi_2+\lambda_4-\lambda_5+ \lambda_6-\lambda_3)+ 2(\chi_1-\chi_2)(-2 \chi_3 + 2 \chi_4 +\lambda_4 +\lambda_5 -\lambda_6-\lambda_3)} =0 \nn\\
&& \nn\\
&-&\frac{(-\mu_1^2+\mu_2^2)(\xi_1- 2 \chi_2 -\lambda_1-\lambda_4-\lambda_6) + \mu_2^2(-2 \chi_1+2\chi_2+\lambda_4-\lambda_5+ \lambda_6-\lambda_3)}{(-\lambda_1-\xi_1-2 \chi_2)(-2 \chi_1+2\chi_2+\lambda_4-\lambda_5+ \lambda_6-\lambda_3)+ 2(\chi_1-\chi_2)(\xi_1- 2 \chi_2 -\lambda_1-\lambda_4-\lambda_6) }    \\
&+& \frac{(-\mu_1^2+\mu_2^2)(-2 \chi_3 + 2 \chi_4 +\lambda_4 +\lambda_5 -\lambda_6-\lambda_3)+(\mu_3^2-\mu_4^2)(-2 \chi_1+2\chi_2+\lambda_4-\lambda_5+ \lambda_6-\lambda_3))}{(\lambda_4+\lambda_5- \lambda_6-\lambda_3)(-2 \chi_1+2\chi_2+\lambda_4-\lambda_5+ \lambda_6-\lambda_3)) + 2(\chi_1-\chi_2)(-2 \chi_3 + 2 \chi_4 +\lambda_4 +\lambda_5 -\lambda_6-\lambda_3)} =0 \,,\nn
\eea
 Also notice that all the relationships discussed can only be approximate since the coupling of the scalar fields to the fermions introduces small corrections.

The vacuum in \eq{vacuumdue} breaks the global symmetry $U(1)_F$
and there seem to be a Goldstone boson in the spectrum. It can be removed by a mass term introduced by hand at an intermediate scale between $v_F$ and $f$. However,
as it is possible to see after fermion will be introduced in the
model, this global symmetry is actually anomalous. This means that the
would-be Goldstone is not part of the physical
spectrum.~\footnote{Alternatively, one can think of the anomaly as
an effective mass for the would-be Goldstone boson that, like the
$\eta^\prime$ of the $U(1)_A$ symmetry of chiral perturbation
theory, becomes massive with a mass of the order of the symmetry
breaking. In our case, this process would make the mass of the
would-be Goldstone boson heavier than those of the other flhiggs.} Also notice that similar anomalies in the gauge groups are automatically compensated by the Goldstone bosons, as it always happens in spontaneously broken gauge theories~\cite{FPPS}; they however reappear above the scale $f$ and may help in the determination of the UV completion of the theory.

In order to give a back-of-the-envelope estimate of  this solution---and to see that it satisfies the requirements outlined in the introduction---it is useful to make a few  approximations. Let us for instance take
\be
\xi_i \simeq \chi_i  \simeq \lambda_1 = \lambda_2 = \lambda_3 =  \chi \quad \quad
\lambda_4 = \lambda_5 = \lambda_6 = \lambda 
\ee
to reduce the number of coefficients.
These approximations are rather natural and do not introduce any fine-tuning.
Accordingly,  the vacuum of the potential in \eq{pot-final} can be given as
 \be
v_F^2= \frac{\mu_1^2-\mu_2^2}{\lambda - \chi} \quad \text{and} \quad
v_W^2=\frac{\mu_1^2-\mu_2^2 + \mu_3^2 -\mu_4^2}{\lambda -\chi}\,.
 \ee
In this simplified case, and by further taking $\lambda -\chi \simeq 1$, $\mu_1^2 - \mu_2^2 \simeq -\mu^2/2$ and $\mu_3^2 -\mu_4^2 \simeq 3 \mu^2/8$ (and $\chi \simeq 1/2$, $\mu_3^2 \simeq 2 \mu^2$ to satisfy \eq{vevwf}) we obtain that
\be
v_W^2 = -\mu^2/4 \quad \text{and} \quad v_F^2 = -\mu^2
\ee
so that for the electroweak vacuum given by its experimental value $v_W = -\mu/2 = 246 \; \text{GeV}$, we find $v_F \simeq$ 500 GeV.  For $f\simeq 1$ TeV,  the parameter $k\equiv v_F^2/f^2$ in the mass textures turns out to be small and of the order of the Cabibbo angle. 

In section \ref{sec:es} we will come back to the vacuum solution in \eq{vacuum} and study it for arbitrary parameters to show the range of masses allowed for the scalar particles as well as for the other states of the model. Before that, we must study the gauge boson sector. As we are about to see, this sector is severely constrained and its consistency with precision electroweak data constrains the possible values of $v_F$ and $g^\prime$ and therefore of $f$ if we want to keep the texture parameter small enough. 

\subsection{Gauge bosons and currents}
\label{sec:gbc}

 After symmetry breaking, the model is described at low-energy by a set of gauge and scalar bosons.  
 We discuss first the  gauge boson sector. Its structure is complicated by the mixing of the standard model gauge bosons to the new states we have introduced.  Our general strategy is to impose that the charged currents of the model coincide with those of the standard model.  This done, we are essentially left with the theory of the standard model with the addition of a massive neutral gauge boson $Z^\prime$ and we must check that its presence affects   the  $\rho$ parameter, the Weinberg angle $\theta_W$, the tree level coefficients of the neutral current   and that the value of the mass of the  $Z^\prime$  are all within the experimental bounds. In this way two of the  free parameters of the model, namely $v_F$ and $g^\prime$ are fixed.
 
At the scale $f$, the symmetry surviving the spontaneous breaking
of $SU(10)$ into $SO(10)$ is  $SU(3)_W\times U(1)_W \times
U(1)_X \times U(1)_F$ and we can write  the effective kinetic
lagrangian for the four scalar triplets $\Phi_i$ as
 \bea
\label{lagfi}
 L_K^\Phi &=&
(D_{\mu}\Phi_1)^{\dag}(D_{\mu}\Phi_1)+(D_{\mu}\Phi_2)^{\dag}(D_{\mu}\Phi_2)+(D_{\mu}^\prime \Phi_3)^{\dag}
(D_{\mu}^\prime \Phi_3)+ (D_{\mu}^\prime \Phi_4)^{\dag}(D_{\mu}^\prime \Phi_4)
\eea
 where the covariant derivatives are given by
\bea
\label{ders}
D_{\mu}&=&\partial_{\mu} + i g W^{a}_{\mu}t^a+ i {g'}x_\Phi B_\mu \nn\\
D_{\mu}^\prime &=&\partial_{\mu} + i g W^{a}_{\mu}t^a+ i {g'}x_\Phi
B_\mu \pm i k\frac{1}{2}X_\mu \, ,
\eea
with, as before in \eq{kinetic},  $W^{a}_{\mu}$ the
gauge bosons of the $SU(3)$ electroweak group, $t^a$ its
generators, $B_\mu$ the gauge boson of the $U(1)$ electroweak
symmetry, $X_\mu$ that of the exotic $U(1)$ gauge symmetry and
$g$, ${g'}$ and $k$ their coupling respectively, while $x_\Phi$ is the
$U(1)$ extended electroweak charge of the triplets
 $\Phi_i$.
Since
the $SU(3) \times U(1)\times U(1)$ gauge symmetry at the low scale
is the diagonal combination surviving the
 spontaneous breaking of $SU(10) $ into $SO(10)$ their couplings are given, respectively, by
 \be
 g^2=\frac{g_1^2 g_2^2}{g_1^2+g_2^2} \, , \quad  {g'}^2=\frac{{g^\prime}^2_1 {g^\prime}^2_2}{{g^\prime_1}^2+ {g^\prime_2}^2}
 \quad \text{and} \quad k^2=\frac{k_1^2 k_2^2}{k_1^2+ k^2_2} \,.
 \ee

When the triplets acquire the vacuum expectation values given by \eq{vacuum}, we are left with nine massive  and one massless gauge boson; this latter  being the photon.

The eight massive gauge bosons can be written as $3$ complex and $3$
real gauge bosons. The lightest complex fields and the lightest real can be identified with the standard model weak gauge bosons
$W$ and $Z$. The remaining complex bosons are new massive charged gauge particles $\tilde W_{1,2}$. The masses of these complex gauge bosons are given by
\be
m^2_W= \frac{1}{4} g^2 v^2_W\, ,  \quad m^2_{\tilde{W}_1}=\frac{1}{2} g^2 v^2_F
\quad \text{and} \quad m^2_{\tilde{W}_2}=\frac{1}{2} g^2( v^2_F +\frac{v^2_W}{2}) \,,
\ee
respectively. The charged $W$ gauge bosons behave exactly like those of the standard model and can be directly identified with them.
Contrary to the heavy gauge bosons in \eq{heavy}, the gauge bosons $\tilde W_{1,2}$ do not mix with $W$ and therefore do not induce additional effective operators in the low-energy theory. Similarly,  the gauge boson of the exotic $U(1)$
gauge symmetry  does not mix and acquires a mass given by
\be
m^2_X= \frac{1}{4}k^2 v^2_F \,.
\ee

The other three real gauge bosons, those associated to the diagonal generators of the
 $SU(3)$, $W_\mu^3$ and $W_\mu^8$, and the gauge boson
of $U(1)_W$, $B$, do mix, and their mass matrix  is given by
\bea
\label{massmat}
M^2_{WB}&=& \left(\begin{array}{ccc}
  g^2 v^2_W/4 &- g^2v^2_W/4\sqrt{3}& -g \tilde{g}^\prime  v^2_W/2   \\
- g^2 v^2_W/4 \sqrt{3}& g^2( v^2_W/12+2 v^2_F/3) &  g \tilde{g}^\prime   ( v^2_W-4 v^2_F)/2\sqrt{3} \\
- g \tilde{g}^\prime  v^2_W/2  &  g \tilde{g}^\prime    ( v^2_W-4 v^2_F)/2\sqrt{3}  &{\tilde{g}^{\prime 2}}   (v^2_W+ 2 v^2_F)  \\
\end{array}\right) \,,
\eea
where in \eq{massmat} ${\tilde{g}^\prime}=g^\prime x_\Phi$.
The $3\times 3$ mixing arises because of the $SU(3)$ weak group we started with and leads to the most characteristic (and constrained) new physics in the model.
 
 One eigenvalue of the matrix $M^2_{WB}$ in \eq{massmat}  is zero and
corresponds to the photon, the other two depend  on the values $v_F$
and $\tilde{g}^\prime$, the lightest mass to be identified  with that of the standard model $Z$, the heaviest with an extra gauge boson $Z^\prime$.

The mixing between $W_3$, $W_8$ and $B$ is delicate since it gives rise to
electric   and neutral currents for the
standard fermions. We  fix the value of $v_F$ and $\tilde{g}^\prime
$ by imposing that the electric and neutral currents in our model
coincide with those of the standard model. In order to analyze
the neutral currents,  consider the orthogonal matrix $U_W$ that
diagonalize $M^2_{WB}$ according to
 \be
 \text{diag} (0,\, M_Z^2,\, M_{Z^\prime}^2) =
U_W^T \,M^2_{WB}\,U_W \,. \label{U}
\ee
Once $g$ and $v_W$ are fixed by their standard model values, the entries of the matrix $U_W$---three of which are independent variables---depend on the parameters $\tilde{g}^\prime$ and $v_F$ that we are going to determine by requiring consistency with the experimental data.

 Consider now the interactions between  a fermion triplet (antitriplet) of $SU(3)_W$, $Q_L\, (Q^*_L)$ , of $U(1)_W$ charge $x_L$
 and two fermion singlets  of $SU(3)_W$, $\psi^{1,2}_R$, of $U(1)_W$ charge $y^{1,2}_R$
  respectively and a fermion singlet  of $SU(3)_W$, $\tilde{\psi}_R$, of $U(1)_W$ charge $\tilde{y}_R$ and of $U(1)_X$ charge $-1/2$,
  with the electroweak gauge bosons, that is we neglect the exotic X-current.
 The first two components  of the left-handed triplet (antitriplet) $Q_L\, (Q^*_L)$,$\psi_L^1$ and $\psi_L^2$ form a $SU(2)_W$
 Standard Model doublet (antidoublet), and when $SU(3)_W \times U(1)_W \times U(1)_X$ is broken into $U(1)_Q$, $\psi^{j}=\psi^{j}_L+\psi^{j}_R $
 has electric charge $Q_{f_{j}}$, with $j=1,2$. At the same time, the third component of the  triplet (antitriplet) $Q_L\, (Q_L^*)$, $\tilde{\psi}_L$ 
 and the exotic $SU(3)_W$ singlet $\tilde{\psi}_R$ give rise to an electric charged  fermion $\tilde{\psi}=\tilde{\psi}_L+\tilde{\psi}_R$, with charged  $Q_{f_{2}}$, where the index $2$ refers tothe second component of the triplet (antitriplet) $Q_L\, (Q_L^*)$.
Dividing the Standard model doublet (antidoublet ) componets,

 The kinetic lagrangian is given by
\be 
\label{fgauget}
L_K^f=
\overline{Q_{L}}\,\gamma \cdot D \, Q_L + \overline{\psi_{R}}\,
\gamma \cdot D \, \psi_R +\overline{\tilde{\psi}_{R}}\,
\gamma \cdot D \, \tilde{\psi}_R   \,, 
\ee
for a triplet and 
\be 
\label{fgauget2}
L_K^f=
\overline{{Q^{*}}}_{L}\,\gamma \cdot D^* \, Q^*_L + \overline{\psi_{R}}\,
\gamma \cdot D \, \psi_R +\overline{\tilde{\psi}_{R}}\,
\gamma \cdot D \, \tilde{\psi}_R   \,, 
\ee
for an antitriplet,
with 
\be 
\label{derf} 
D_\mu  =
\partial_\mu +  i g W^{a}_{\mu}t^a+ i {g'} x^{j}_{L,R} B_\mu \,,
\ee
where in  \eq{derf} have been used the same notations as in
\eq{ders}. Consider only the terms in \eqs{fgauget}{fgauget2} that give rise
to the electromagnetic and the neutral current  for all the  fermions, that is
 \bea
 \label{ncvec}
 \mathcal{L}_K^{f}&=& \overline{\psi_L^j}\gamma^\mu \,\Big(\partial_\mu + i
 g\,T_{3\,f_{j}}\,W_\mu^3 + i\,g\, \frac{p}{2 \sqrt{3}}\,W_\mu^8 +
 i\,g^\prime\, x_L B_\mu \Big)\psi_L^j \nn\\
&+&
 \overline{\psi_R^j}\gamma^\mu \,\Big(\partial_\mu+i\,g^\prime\, x_R^j B_\mu \Big)\psi_R^j
 \nn\\
&+& \overline{\tilde{\psi}_L}\,\gamma^\mu \,\Big(\partial_\mu - i\,g\, \frac{ p}{\sqrt{3}}\,W_\mu^8 +
 i\,g^\prime\, x_L B_\mu \Big)\tilde{\psi}_L + \overline{\tilde{\psi}_R}\gamma^\mu \,\Big(\partial_\mu+i\,g^\prime\, \tilde{x}_R B_\mu \Big)\tilde{\psi}_R
 \,,
 \eea
where we have explicited the standard model doublet (antidoublet) components $\psi_L^{1,2}$ and the exotic fermion $\tilde{\psi}_L$ and where $p$ is equal to $1$ or $-1$ for the left handed fermion coming from a triplet or an antitriplet respectively. 

The gauge bosons $W^3_\mu$, $W^8_\mu$ and $B_\mu$ mix through the
$U_W$ of \eq{U} giving the photon, $A_\mu$ the gauge boson $Z_\mu$
and an heavy Z-type gauge boson, ${Z'}_\mu$, in particular we have
\bea 
\label{nuovigauge} 
\left(\begin{array}{c}
W_3\\
W_8\\
B\\\end{array}\right)&=& U_W \, \left(\begin{array}{c}
A\\
Z\\
{Z'}\\\end{array}\right) \,. 
\eea 
Substituting the expressions
coming from \eq{nuovigauge} in \eq{ncvec} we can write the
electric, the neutral and the extra neutral
 currents using the parametrization  given in \cite{PDG}
\bea 
\label{currents}
\mathcal{L}^\prime &=& - e Q_j \overline{\psi^j}\gamma^\mu
\psi^j A_\mu - \frac{e}{2 s_W c_W}\,\Big( 1+ \frac{\alpha T}{2} \Big)
\overline{\psi^j}\gamma^\mu \big(g_V^j - g_A^j \gamma^5
\big)\psi^j\,
Z_\mu \nn\\
&-&  \frac{e}{2 s_W c_W}\,\overline{\psi^j}\gamma^\mu
\big(\tilde{h}_V^j - \tilde{h}_A^j \gamma^5 \big)\psi^j\,
Z^\prime_\mu \nn\\
&-& e  Q_2 \overline{\tilde{\psi}}\gamma^\mu
\tilde{\psi} A_\mu - \frac{e}{2 s_W c_W}\,\overline{\tilde{\psi}}\gamma^\mu
\big(g_V^3 - g_A^3 \gamma^5 \big)\overline{\psi} Z_\mu \nn \\
\,
&-& \frac{e}{2 s_W c_W}\,\overline{\tilde{\psi}}\gamma^\mu
\big(\tilde{h}_V^3 - \tilde{h}_A^3 \gamma^5 \big)\overline{\psi}Z^\prime_\mu \,, \eea
where $s_W$ and $c_W$ are  sine and  cosine of the Weinberg angle
$\theta_W$, $T$ is one of the oblique parameters, 
 \be 
 g_{V,A}^j =  g_{V,A}^{j\,SM}+
\tilde{g}_{V,A}^j \,,
\ee 
with 
\be 
g_{V}^{j\,SM} = T_{3_{j}}- 2
Q_j s_*^2 \quad \text{and} \quad g_A^{j\,SM} = T_{3_{j}}\,,
\ee 
and $g_{V,A}^3$ and $\tilde{h}_{V,A}^3$ are related to the  neutral currents of the exotic fermions $\tilde{\psi}$ that, as we shall show in section \ref{sec:itf} below are only weakly coupled to the standard model states.
The coefficients $\tilde g_{V,A}$ contain the deviation from the standard model, $\tilde{h}_{V,A}$ the strength of the coupling of $Z^\prime$ to the standard model fermions.

First of all, the entries of the orthogonal matrix $U_W$ have to satisfy
the following conditions in order to have in \eq{currents} the
correct electric charge current for the standard and exotic fermions:
\bea 
\label{condcor}
U_{W_{1\,1}}&=& s_W
\nn\\
\frac{U_{W_{2\,1}}}{U_{W_{1\,1}}} &=& \frac{1}{\sqrt{3}} \nn \\
\frac{1}{U_{W_{1\,1}}} \Big( U_{W_{2\,1}}\frac{p}{2 \sqrt{3}}
+x_L \frac{g^\prime}{g}U_{W_{3\,1}}  \Big)&=& x_L^{SM} \nn\\
\frac{g^\prime}{g} x_R^j\, \frac{U_{W_{3\,1}}}{U_{W_{1\,1}}}  &=& Q_j \nn\\
 \frac{g^\prime}{g} \tilde{x}_R\, \frac{U_{W_{3\,1}}}{U_{W_{1\,1}}}  &=& Q_2 \,,
\eea
where $x_L$ and $x_R^j$ are the extra $U(1)_W$ fermion charges,
while $x_L^{SM}$ is the $U(1)_W$ standard model charge of the
electroweak doublet $\psi_L$. 
The conditions of \eq{condcor} together with
\be
\label{condcorfi}
\frac{1}{U_{W_{1\,1}}} \Big( U_{W_{2\,1}}\frac{1}{\sqrt{12}}
+ \frac{\tilde{g}^\prime}{g} U_{W_{3\,1}}  \Big) = 0 \,,
\ee
that follows by inserting \eq{condcor}  in \eq{lagfi} and imposing zero electric charge for  the    second and the third components of the triplets $\Phi_i$ completely determines all the independent parameters of the matrix $U_W$.

Since the mass matrix of \eq{massmat} depends on $\tilde{g}^\prime= g^\prime x_\Phi$ the other three conditions give us the values of  $x_L / x_\Phi $ , $x^j_R/ x_\Phi  $ and $\tilde{x}_R/ x_\Phi  $, that is the fermion charges in units of the triplet $\Phi_i$ charges, with the further constrain on $U_{W_{31}}$ of giving rational numbers for the charges.
 
By equating now the neutral currents, we obtain that 
\bea 
(1+\frac{\alpha T}{2}) &=& c_W
\,U_{W_{1\,2}}
\Big(1-\frac{U_{W_{3\,2}}}{U_{W_{3\,1}}}\frac{U_{W_{1\,1}}}{U_{W_{1\,2}}}
\Big)
\nn\\
(1+\frac{\alpha T}{2}) \; s_*^2 &=& - c_W
\frac{U_{W_{3\,2}}U_{W_{1\,1}} }{U_{W_{3\,1}}} 
 \nn\\
(1+\frac{\alpha T}{2})\; \tilde{g}_{V}^j=(1+\frac{\alpha T}{2}) \; \tilde{g}_{A}^j &=& p \, c_W \,\frac{U_{W_{2\,2}}}{2 \sqrt{3}}
\Big(1-\frac{U_{W_{3\,2}}}{U_{W_{3\,1}}}\frac{U_{W_{2\,1}}}{U_{W_{2\,2}}}\Big)
\nn\\
\tilde{h}_{V}^j &=& T_{3_{j}} U_{W_{1\,3}}\,
\Big(1-\frac{U_{W_{3\,3}}}{U_{W_{3\,1}}}\frac{U_{W_{1\,1}}}{U_{W_{1\,3}}}
\Big)+ 2 \frac{U_{W_{3\,3}}}{U_{W_{3\,1}}}U_{W_{1\,1}} Q_j
+p\, \frac{U_{W_{2\,3}}}{\sqrt{12}}\Big(1-\frac{U_{W_{3\,3}}}{U_{W_{3\,1}}}\frac{U_{W_{2\,1}}}{U_{W_{2\,3}}}\Big)
\nn\\
\tilde{h}_{A}^j &=& T_{3_{j}} U_{W_{1\,3}}\,
\Big(1-\frac{U_{W_{3\,3}}}{U_{W_{3\,1}}}\frac{U_{W_{1\,1}}}{U_{W_{1\,3}}}
\Big)
+p\,\frac{U_{W_{2\,3}}}{2 \sqrt{3}}\Big(1-\frac{U_{W_{3\,3}}}{U_{W_{3\,1}}}\frac{U_{W_{2\,1}}}{U_{W_{2\,3}}}\Big)
\,. \label{sTgf}
\eea 

A more complete analysis would require that also the corrections arising from the effective operators induced by the heavy gauge bosons in \eq{heavy} be included. Thy are important because they violate the $SU(2)$ custodial symmetry of the standard model. They  affect the relationships in \eq{sTgf} to ${ O} (v^2_W/f^2)$ and, in the littlest Higgs model of ref.~\cite{littlest}, force the scale $f$ to be above 4 TeV~\cite{littlehiggs2}. As already mentioned, these constrains can be lessen by introducing an additional discrete symmetry~\cite{Cheng}. Notice that the overall fit of these  corrections against the experimental electroweak data can in principle be improved by the presence in the flhiggs model of the additional parameters in \eq{sTgf} thus lowering the scale $f$ with respect to that found in the case of the littlest Higgs model. 

As we said in section \ref{sec:mb}, we are more interested in the consistency of the framework than in its detailed realization and therefore we neglect, in this work, these ${ O} (v^2_W/f^2)$ corrections and consider \eq{sTgf} as it stands. 

The parameters and coefficients in \eq{sTgf} are constrained by precision measuraments of neutral currents in low-energy observables like atomic parity violation in atoms and neutrino-hadron scattering.
The mass of the $Z^\prime$ gauge boson is bounded by data on Drell-Yan production (with subsequent decay into charged leptons) in $p\bar p$ scattering to be larger than 690 GeV~\cite{Abe} and this constrain must be included as well. 
We require that deviation in the $\rho$-parameter
\be
\rho = 1+ \alpha T
\ee
and the Weinberg angle be within $10^{-3}$ while the $\tilde g$ coefficients in \eq{sTgf} be less than $10^{-2}$. This choice put these deviations in the  tree-level parameters in the ball park of standard-model radiative corrections. 

The importance of these constrains resides in their fixing the values of the free parameters $v_F$ and $\tilde{g}^\prime= g^\prime x_\Phi $. 
The bound on the $\rho$ parameter essentially fixes the effective gauge coupling\be
\tilde{g}^\prime \simeq  0.13 \, .
\ee
For simplicity we take $x_\Phi =1$ so we have $\tilde{g}^\prime =g^\prime $.

Once $\tilde{g}^\prime$ has been fixed to this value, the bound  on the mass of $Z^\prime$ requires
 \be
v_F \gtap 1260 \quad \text{GeV}\, .
\ee 
 We would like to have $v_F$ as close to $v_W$ as possible but the phenomenological constrains force it to a higher scale.
 
The rather large value we must take for $v_F$ does   imply unfortunately that some amount of fine tuning in the parameters of the potential in \eq{pot} is present. If we go back to our back-of-the-envelope estimate in section \ref{sec:vev}, we see that while there we had $v_f \simeq 2 v_W$ with  no fine tuning (that is, the coefficient were chosen with a tuning of one out of four or 25\%)  on the values of the $\mu_i$ coefficients,  we now must have $v_f\simeq 4 v_W$ that can be obtained by taking, for instance, $\mu_1^2 - \mu_2^2 \simeq -\mu^2/2$ and $\mu_3^2 -\mu_4^2 \simeq 24 \mu^2/ 50$
that means 1 out of 25, that is a  fine tuning of 4\%.  The actual fine-tuning in the model is however less than this because of the larger  number of parameters involved and roughly of 10\% or less.

\subsection{The scalar sector and the lightest flhiggs boson}

 We now turn to the scalar  sector of the model.
 The number of scalar bosons can readily be computed:
 the number of degrees of freedom of  $4$ complex triplets is $24$. Of these $9$ are eaten by the gauge fields, while $1$---the would-be Goldstone boson of the spontaneous breaking of the $U(1)_F$ global symmetry---is eliminated, after introducing the fermions in the model,  by the anomaly. Therefore, the scalar sector contains  $24-10= 14$  massive fields. To describe them,
 we  parametrize the $\Phi_{i}$
  triplets with respect to these fifteen  fields as follows
\bea
\Phi_1&=&\left(\begin{array}{c}
  u^\rho_{11}\rho_1 \\
 (v_W + u^\delta_{1i}\delta_{i}+u^\varphi_{1j}\varphi_j )/2\\
  (v_F+ u^\delta_{2i}\delta_{i}+u^\varphi_{2j}\varphi_j)/2\\ \end{array}
  \right)\nn\\
  \Phi_2&=&\left(\begin{array}{c}
  u^\rho_{21}\rho_1 \\
 (v_W + u^\delta_{3i}\delta_{i}+u^\varphi_{3j}\varphi_j )/2\\
  (v_F+ u^\delta_{4i}\delta_{i}+u^\varphi_{4j}\varphi_j)/2 \\\end{array}
  \right)\nn\\
   \Phi_3&=&\left(\begin{array}{c}
  u^\rho_{32}\rho_2 \\
 (u^\delta_{5i}\delta_{i}+u^\varphi_{5j}\varphi_j )/2\\
  (v_F+ u^\delta_{6i}\delta_{i}+u^\varphi_{6j}\varphi_j)/2\\ \end{array}
  \right)\nn\\
  \Phi_4&=&\left(\begin{array}{c}
  u^\rho_{42}\rho_2 \\
 (u^\delta_{7i}\delta_{i}+u^\varphi_{7j}\varphi_j )/2\\
  (v_F+ u^\delta_{8i}\delta_{i}+u^\varphi_{8j}\varphi_j)/2\\ \end{array}
  \right) \,.
  \eea
with $i=1,..,4$ and $j=1,..,7$ and where
$u^{\rho,\delta,\varphi}_{ij}$ are the entries of the unitary
 matrix which diagonalizes the mass matrix defined as
 \be
 M^2_{\Phi_{ij}} = \left. \frac{\partial^2 V[\Phi]}{\partial \Phi_i \partial
 \Phi_j} \right|_{\Phi=\vev{\Phi}} \, ,
 \ee
 and can be written in terms of the coefficients of the effective potential.

The scalar fields $\rho_k$, $\delta_i$ and $\varphi_j$ are the Higgs-like components of the
flhiggs fields and the most interesting experimental signature of the model.

 The fields $\rho_{1,2}$ are electrically charged, their masses given respectively by
by
\be
m^2_{\rho_{1}} = 4 \xi_1(v_F^2-v_W^2) \quad \mbox{and} \quad
m^2_{\rho_{2}} = (\xi_1- \xi_2) v_F^2 -  \xi_1 v_W^2 \,. \label{mass1}
\ee
We shall call the lightest of the two $h^\pm$.

 The masses of the neutral $\delta_i$ fields are
given by
\bea
m^2_{\delta_{1}}&=& 4 \xi_1 (v_F^2-v_W^2)\nn\\
m^2_{\delta_{3}}&=& 2 \xi_1  v_F^2 - \xi_1 v_W^2 \left( 1 - \frac{ v_W^2}{v_F^2} \right) \nn\\
m^2_{\delta_{4}}&=&(\xi_1- \xi_2) v_F^2 -  \xi_1 \frac{ v_W^2}{v_F^2} v_W^2 \, .\label{mass2}
\eea
The missing  $\delta_2$ field, that in the diagonalization  appears as a massless state, is the would-be Goldstone boson eliminated by the anomaly.

The masses of the fields $\varphi_j$ are obtained by
diagonalization of the remaining sub-matrix. This sub-matrix written in terms of the vacua and the coefficients of the potential has a cumbersome  form that is not particularly inspiring and that we do not include.  We do not have an exact diagonalization for it but it must contain the lightest neutral scalar boson that we call $h^0$. This can be understood by thinking at one single flhiggs triplet for which the $\delta_i$ part would correspond to the imaginary component and the $\varphi_j$ to the real part and therefore  Higgs-like component.

We study the scalar sector spectrum numerically in section \ref{sec:es} to obtain an estimate of the allowed values for $m_{h^0}$ and $m_{h^\pm}$ for  arbitrary ${O}(1)$ coefficients in the potential and $v_F$  fixed to the values determined in the previous section.

 \subsection{The flavorless limit}
\label{sec:tfl}

In the limit in which we factorize out the flavor part by taking $v_F=f$, the model has the littlest Higgs model of ref.~\cite{littlest} embedded inside. We can identify within the global symmetry $SU(10)$ a reduced symmetry  $SU(5)$.
The two flhiggs fields $\Phi_{3,4}$ decouple from this subsector that feels  no  $U(1)_X$ symmetry.
In the notation  of \cite{littlest}, the flhiggs fields
 $\Phi_{1,2}$ go into the Higgs boson $h$ while the fields $z_{ij}$ go into the  weak triplet field $\phi$. Clearly, all the Yukawa coupling of the next section become trivial and the fermion masses degenerate if we take the Yukawa coefficients to be all of $\mathcal{O}(1)$.

\section{Introducing the fermions} 
\label{sec:itf}

According to the rules of the little-Higgs mechanism, the coupling of the fermions to the flhiggs must proceed by preserving enough symmetry not to give rise to 1-loop quadratic divergent contributions to their masses. This means that for every fermion with an Yukawa coupling of ${O}(1)$ we must introduce one (or even more) state to cancel the divergent diagram. This procedure brings in two more sets of fermions, one for the standard model fermions with large Yukawa couplings and one for the exotic states we  introduced to complete the $SU(3)$ triplets.

Even though the introduction of new fermions seems to lead us to a structure of Baroque richness, notice that these states nicely fall into the fundamental representations of $SU(10)$ giving a natural structure to the Yukawa interactions in terms of the larger symmetry group that can be written in general, and by neglecting for the moment the flavor group, as
\be
\mathcal{L}_Y \simeq \lambda_1 \bar {\cal  X X} + \lambda_2  \bar {\cal  X} \Sigma {\cal X}
 \ee
where $\cal X$ is a decuplet of fermions in the fundamental representation of $SU(10)$.  

The Yukawa lagrangians at the scale $f$ is obtained by writing the $SU(3)_W \times U(1)_X \times U(1)_F$ invariant terms involving the four triplets $\Phi_1$,$\Phi_2$ ,$\Phi_3$ and $\Phi_4$, and the fermions. Standard model left-handed doublets are members of an $SU(3)_W$ triplet, the third component being an exotic fermion.

To help the reader in keeping track of the various terms, Tables \ref{tabf}--\ref{tabnu} contain the representations and the charge assignments with respect to the exotic, the flavor and the electroweak groups of all fermions and flhiggs bosons, the latter having been determined solving \eq{condcor} and \eq{condcorfi}.
\begin{table}[ht]
\begin{center}
\caption{Representations and charges assignments for the flhiggs bosons.\label{tabf}} 
 \vspace{0.2cm}
\begin{tabular}{|c|c|c|c|c|}
\hline & $U(1)_X$\quad&$U(1)_F$\quad &$SU(3)_W$ \quad& $U(1)_W$\\
\hline \hline
 $ \Phi_1$ & 0&$1/2$&$3$  & $1$\\
\hline
 $ \Phi_2$ &$0$&$-1/2$& $3$  & $1$\\
\hline
 $\Phi_3$ & 1/2&$0$&$3$  & $1$\\
\hline
 $\Phi_4$ &$-1/2$&$0$& $3$  & $1$\\
\hline
\end{tabular}
\end{center}
\end{table}
\begin{table}[ht]
\begin{center}
\caption{Representations and charges assignments for the quarks. Different families run over the index $i$; they differ only for the flavor charges that are written as $(q_1,q_ 2, q_3)$ for, respectively, the first, second and third family. $U(1)_W$ charges are  determined by the data constrains (see text of main body).\label{tabq}} 
\vspace{0.2cm}
\begin{tabular}{|c|c|c|c|c|c|}
\hline & &$U(1)_X$\quad&$U(1)_F$\quad&$SU(3)_W$\quad \quad&
$U(1)_W$\quad \\
\hline \hline
$\mathcal{Q}^i_L=\left(\begin{array}{c}Q^i_L \\0\\\tilde{Q}_L^i\\ U_L^i\\ \tilde{U}_L^i\\ 0\\ \end{array}\right) $ &
\begin{tabular}{c} $ Q^i_L=\left(\begin{array}{c} d_L\\ u_L\\ \tilde{u}_L\\
\end{array} \right)$ \\\hline   
$ \tilde{Q}^i_L=\left(\begin{array}{c} m^i_L\\ n^i_L\\ \tilde{n}^i_L\\
\end{array} \right)$\\\hline
$U^{i}_L $\\  \hline $\tilde{U}^{i}_L $  \end{tabular} &
\begin{tabular}{c}\,$\begin{array}{c}\, \\0 \\ \,\\
\end{array}$ \\ \hline $\begin{array}{c} \,\\ 0\\ \,\\
\end{array}$\\ \hline $0$\\ \hline $1/2$\\ \end {tabular} &
\begin{tabular}{c}$\begin{array}{c}\, \\(9/2, 7/2, 3/2) \\ \,\\
\end{array}$ \\ \hline  $\begin{array}{c} \,\\ (9/2, 7/2, 3/2)\\ \,\\
\end{array}$ \\ \hline $(4, 3, 1)$\\ \hline $(9/2, 7/2, 3/2)$\\ \end {tabular} & \begin{tabular}{c}$\begin{array}{c}\, \\ \overline{3}\\ \,\\
\end{array}$ \\ \hline $\begin{array}{c} \,\\ 3\\ \,\\
\end{array}$ \\ \hline $1$\\ \hline $1$\\ \end {tabular} &\begin{tabular}{c}$\begin{array}{c}\, \\ 1\\ \,\\
\end{array}$ \\ \hline $\begin{array}{c} \,\\ 3\\ \,\\
\end{array}$ \\ \hline $2$\\ \hline $2$\\ \end {tabular} \\
\hline
$\mathcal{U}^{i^{c}}_L=\left(\begin{array}{c}0_\alpha \\{u'}^{i^{c}}_L\\0_\alpha\\ 0\\ 0\\0\\ \end{array}\right)$&${u'}^{i^{c}}_L$ &  $0$&$(3, 1, 0)$& $1$  & $-2$\\
\hline $\tilde{\mathcal{U}}^{i^{c}}_L=\left(\begin{array}{c}0_\alpha
\\0\\0_\alpha\\ 0\\ 0\\{\tilde{u'}}^{i^{c}}_L\\
\end{array}\right)$&${\tilde{u'}}^{i^{c}}_L$ &  $-1/2$&$(-9/2, -7/2, -3/2)$& $1$
& $-2$\\\hline
$\tilde{Q}^{i^{c}}_L$ &$\left(\begin{array}{c} x^{c^{i}}\\ y^{c^{i}}\\ \tilde{y}^{c^{i}}\\ \end{array} \right)$& $0$ & $(-9/2, -7/2, -3/2)$& $\overline{3}$&$-3$ \\
\hline
&${u''}^{i^{c}}$& $0$& $(-3, -2, 0)$&$1$&$-2$ \\
\hline
&$\tilde{u''}^{i^{c}}$& $-1/2$& $(-9/2, -7/2,-3/2) $&$1$&$-2$ \\
\hline
&$d_L^{i^{c}}$ & $0$& $(7/2, 5/2, 5/2)$&$1$&$1$ \\
\hline
\end{tabular}
\end{center}
\end{table}

\subsection{Quarks}

In order to avoid  large quadratic corrections to the flhiggs masses  induced by divergent one-loop contributions from the heaviest fermions present in the model, that is the top and the exotic quarks (and leptons) that complete the electroweak triplets,   we introduce for each family  a  number of colored Weyl fremions, both triplets of the $SU(3)$ electroweak gauge group and singlets. Their charges are all summarized in  \tab{tabq}. The number of multiplets and singlets introduced is the smallest number that permit us to write a quark Yukawa lagangian composed by terms that singularly preserve enough symmetry in order to keep the four triplets $\Phi_1$, $\Phi_2$, $\Phi_3$ and $\Phi_4$  massless. In this way  quadratic divergent contributions to the flhiggs masses arise only at two-loops.

The Yukawa lagrangian for the quarks is given by
\bea
\label{quarkphi}
\mathcal{L}^q_Y & = & \lambda^{u\,1}_{ab}\,f\,\mathcal{U}^{a^{c}}_L\,\Sigma \, \mathcal{Q}^b_L
\Big(\Sigma_{4,4}\Big)^{|y^b_{Q}+ y^a_{\mathcal{U}}|} + \lambda^{u\,2}_{ab}\,f\, {u''}^{a^{c}}_L U_L^b
 \Big(\Sigma_{4,4}\Big)^{|1+ y^b_{Q}- y^a_{Q}|} \nn \\
&+& \tilde{\lambda}^{u\,1}_{ab}\,f\,\tilde{\mathcal{U}}^{a^{c}}_L\,\Sigma \, \mathcal{Q}^b_L
\Big(\Sigma_{4,4}\Big)^{|y^b_{\tilde{U}}- y^a_{\tilde{U}}|} + \tilde{\lambda}^{u\,2}_{ab}\,f\,
{\tilde{u''}}^{a^{c}}_L \tilde{U}_L^b  \Big(\Sigma_{4,4}\Big)^{|y^b_{\tilde{U}}- y^a_{\tilde{U}}|}\nn\\
&+&\eta^u_{ab}\, f \, \tilde{Q}^{a^{c}}_L \tilde{Q}_L^b
\Big(\Sigma_{4,4}\Big)^{|y^b_{Q}- y^a_{Q}|}+ \lambda^d_{ab}
d^{a^{c}}_L \Big( \epsilon{ijk}Q_{L_{i}}^b \Sigma_{4,4+j}
\Sigma_{8,4+i}\Big) \Big(\Sigma_{4,4}\Big)^{|y^b_{Q}+ y^a_{d}|}
 \eea
 with $a=1,2,3$ and
$d_L^{{1,2,3}^c}=d_L^c,s^c_L,b^c_L$. $y_Q^{a}$ are the flavor
charges of the $Q_L^a$ and $\tilde{Q}_L^a$ triplets,  now members of
the $\mathcal{Q}^a$ multiplets defined in
\tab{tabq}, $y_{\mathcal{U}}^{a}$ that  of the
$\mathcal{U}^a$ multiplets and $y_d^a$ the flavor charges of the
Weyl fermions $d_L^{a^{c}}$. In \eq{quarkphi} the terms with the coefficients  $\lambda^{u\,1}_{ab}$ preserve an $SU(8)$ subgroup of the approximate global symmetry $SU(10)$ while the terms with the coefficients    $\lambda^{u\,2 }_{ab}$ break it but preserve  an $SU(9)$ subgroup of $SU(10)$. Analogously, the terms with coefficients  $\tilde{\lambda}^{u\,1 }_{ab}$ and  $\tilde{\lambda}^{u\,2 }_{ab}$ preserve different subgroups of $SU(10)$ making possible the protection of the flhiggs masses through the collective symmetry breaking mechanism. 

The $(4,\, 4)$ component of the sigma model field $\Sigma$
that appears in \eq{quarkphi} is the only $SU(8)$ and $SU(9)$ singlet and therefore the only possible field we can include to balance the flavor charges to make \eq{quarkphi} invariant.

Expanding the $\Sigma$ and keeping
only the terms involving the $\Phi_i$ in \eq{quarkphi} yields
\bea
\label{quarkphiex}
\mathcal{L}^q_Y &=& \lambda^{u\,1}_{ab}\,f\, \Big[{u'}^{a^{c}}_L U^b_L \big(1-\frac{\Phi_1^\dag \Phi_1}{2 f^2} -\frac{\Phi_2^\dag \Phi_2}{2 f^2}\big)+{u'}^{a^{c}}_L \big(i\frac{\Phi_2^T}{f}Q_L^b + i\frac{\Phi_1^\dag}{f}\tilde{Q}_L^b  \big)   \Big]\Big(\frac{\Phi_1^\dag \Phi_2}{f^2}\Big)^{|y^b_{Q}+ y^a_{\mathcal{U}}|}\nn\\
& +& \lambda^{u\,2}_{ab}\,f\,{u''}^{a^{c}}_L U_L^b  \Big(\frac{\Phi_1^\dag \Phi_2}{f^2}\Big)^{|1+ y^b_{Q}- y^a_{Q}|} \nn \\
&+& \tilde{\lambda}^{u\,1}_{ab}\,f\,\Big[{\tilde{u'}}^{a^{c}}_L \tilde{U}^b_L \big(1-\frac{\Phi_3^\dag \Phi_3}{2 f^2} -\frac{\Phi_4^\dag \Phi_4}{2 f^2}\big)+{\tilde{u'}}^{a^{c}}_L \big(i\frac{\Phi_3^T}{f}Q_L^b + i\frac{\Phi_4^\dag}{f}\tilde{Q}_L^b  \big)\Big] \Big(\frac{\Phi_1^\dag \Phi_2}{f^2}\Big)^{|y^b_{\tilde{U}}- y^a_{\tilde{U}}|}\nn\\
& +& \tilde{\lambda}^{u\,2}_{ab}\,f\, {\tilde{u''}}^{a^{c}}_L \tilde{U}_L^b  \Big(\frac{\Phi_1^\dag \Phi_2}{f^2}\Big)^{|y^b_{\tilde{U}}- y^a_{\tilde{U}}|}\nn\\
&+&\eta^u_{ab}\, f \, \tilde{Q}^{a^{c}}_L \tilde{Q}_L^b \Big(\frac{\Phi_1^\dag \Phi_2}{f^2}\Big)^{|y^b_{Q}- y^a_{Q}|}+ \lambda^d_{ab} d^{a^{c}}_L \Big( \epsilon{ijk}Q_{L_{i}}^b \frac{\Phi_1^*}{f} \frac{\Phi_2^*}{f} \Big) \Big(\frac{\Phi_1^\dag \Phi_2}{f^2}\Big)^{|y^b_{Q}+ y^a_{d}|} \, .
\eea

After the symmetry breaking (\ref{vacuumdue}) that  for convenience we rewrite here
\be
\vev{\Phi_1}= \left(\begin{array}{c}
  0 \\
 v_W /2\\
  v_{F}/2 \\
\end{array}\right)
\quad \quad
\vev{\Phi_2}= \left(\begin{array}{c}
  0 \\
 v_W /2\\
 - v_{F}/2  \\
\end{array}\right) \quad \quad
\vev{\Phi_3}= \left(\begin{array}{c}
  0 \\
 0\\
 v_{F}/2  \\
\end{array}\right)\quad \quad
\vev{\Phi_4}= \left(\begin{array}{c}
  0 \\
 0\\
 v_{F}/2  \\
\end{array}\right) \label{ssb}
\ee
the Yukawa lagrangian  \eq{quarkphiex} at the leading order becomes
\bea
\label{quarkphivev}
\mathcal{L}^q_Y &=&
\lambda^{u\,1}_{ab}\,f\, \Big[{u'}^{a^{c}}_L U^b_L +{u'}^{a^{c}}_L
\big(\frac{v_W}{f} u^b_L- \frac{v_F}{f} \tilde{u}^b_L +
\frac{v_W}{f} n^b_L+ \frac{v_F}{f} \tilde{n}^b_L \big)\Big]
\big(-k \big)^{|n^b_{Q}+ n^a_{\mathcal{U}}|}+
\lambda^{u\,2}_{ab}\,f\,{u''}^{a^{c}}_L U_L^b
 \big(-k \big)^{|1+ n^b_{Q}- n^a_{Q}|} \nn \\
 &+& \tilde{\lambda}^{u\,1}_{ab}\,f\,\Big[{\tilde{u'}}^{a^{c}}_L \tilde{U}^b_L +{\tilde{u'}}^{a^{c}}_L
 \big(\frac{v_F}{f} \tilde{u}_L^b + \frac{v_F}{f} \tilde{n}^b_L  \big)\Big] \big( - k \big)^{|n^b_{\tilde{U}}- n^a_{\tilde{U}}|}+ \tilde{\lambda}^{u\,2}_{ab}\,f\, {\tilde{u''}}^{a^{c}}_L \tilde{U}_L^b  \big(-k \big)^{|n^b_{\tilde{U}}- n^a_{\tilde{U}}|}\nn\\
&+&\eta^u_{ab}\, f \, (m^{a^{c}}_L m^b_L +n^{a^{c}}_L n^b_L +\tilde{n}^{a^{c}}_L \tilde{n}^b_L ) \big(-k \big)^{|n^b_{Q}- n^a_{Q}|}+ \lambda^d_{ab} d^{a^{c}}_L d^b_L 2 v_W \epsilon \big(-k\big)^{|n^b_{Q}+ n^a_{d}|} \, ,
\eea
where $k=v^2_F/f^2$ is the parameter in terms of which we write the mass textures.

From the lagrangian in \eq{quarkphivev}, we can  read off  the mass matrices for the quarks (see also \eq{quarkphilight} below). These matrices and their textures are discussed in section \ref{sec:mmm}. 

\subsection{Collective breaking in the up-quark sector and decoupling of the exotic fermions}

Let us now pause for a moment and show how the collective breaking mechanism works in  preventing 1-loop quadratically divergent corrections to the flhiggs masses.
Consider  only the terms of the type ``up'' components of the third family
\bea
\mathcal{L}^Y_{top} &=&\lambda^{u\,1}_{33}\,f\, \Big[{t'}^{c}_L T_L +{t'}^{c}_L \big(\frac{v_W}{f} t_L- \frac{v_F }{f}
\tilde{t}_L + \frac{v_W}{f} n^3_L+ \frac{v_F}{f} \tilde{n}^3_L \big)   \Big] \big(-k \big)+
\lambda^{u\,2}_{33}\,f\,{t''}^{c}_L T_L  \big(-k \big) \nn \\
&+& \tilde{\lambda}^{u\,1}_{33}\,f\,\Big[{\tilde{t'}}^c_L \tilde{T}_L +{\tilde{t'}}^c_L
\big(\frac{v_F}{f} \tilde{t}_L + \frac{v_F}{f} \tilde{n}^3_L  \big)\Big] +
\tilde{\lambda}^{u\,2}_{33}\,f\, {\tilde{t''}}^c_L \tilde{T}_L \nn\\
&+&\eta^u_{ab}\, f \, (m^{3^{c}}_L m^3_L +n^{3^{c}}_L n^3_L
+\tilde{n}^{3^{c}}_L \tilde{n}^3_L ) \,. 
\eea 
We see that
${t'}_L^c$ and  ${t''}_L^c$ mix into a heavy and light
combination, the latter being the standard top quark $t_L^c$. Since the
mixing is given by 
\bea
\label{top}
t_L^c &= & \frac{\lambda^{u\,1}_{33}}{\sqrt{(\lambda^{u\,1}_{33})^2+(\lambda^{u\,2}_{33})^2}} {t''}_L^c - \frac{\lambda^{u\,2}_{33}}{\sqrt{(\lambda^{u\,1}_{33})^2+(\lambda^{u\,2}_{33})^2}} {t'}^c \nn\\
\hat{t}_L^c &= &
\frac{\lambda^{u\,2}_{33}}{\sqrt{(\lambda^{u\,1}_{33})^2+(\lambda^{u\,2}_{33})^2}}
{t''}_L^c +
\frac{\lambda^{u\,1}_{33}}{\sqrt{(\lambda^{u\,1}_{33})^2+(\lambda^{u\,2}_{33})^2}}
{t'}_L^c \,, 
\eea 
the top Yukawa coupling is 
 \be
\lambda^u_{33} =
\frac{\lambda^{u\,1}_{33}\lambda^{u\,2}_{33}}{\sqrt{(\lambda^{u\,1}_{33})^2+(\lambda^{u\,2}_{33})^2}}
\,. 
\ee

Similarly, the exotic $\tilde{t'}^c_L$ and $\tilde{t''}^c_L$
mix into a heavy and a light combination, giving rise to the
exotic quarks $\tilde{t}^c_L$ and $\tilde{\hat{t}}^c_L$: 
\bea
\tilde{t}_L^c &= &
\frac{\tilde{\lambda}^{u\,1}_{33}}{\sqrt{(\tilde{\lambda}^{u\,1}_{33})^2+
(\tilde{\lambda}^{u\,2}_{33})^2}} \tilde{t''}_L^c -
\frac{\tilde{\lambda}^{u\,2}_{33}}
{\sqrt{(\tilde{\lambda}^{u\,1}_{33})^2+(\tilde{\lambda}^{u\,2}_{33})^2}} \tilde{t'}_L^c \nn\\
\tilde{\hat{t}}_L^c &= &
\frac{\tilde{\lambda}^{u\,2}_{33}}{\sqrt{(\tilde{\lambda}^{u\,1}_{33})^2+(\tilde{\lambda}^{u\,2}_{33})^2}}
\tilde{t''}_L^c +
\frac{\tilde{\lambda}^{u\,1}_{33}}{\sqrt{(\tilde{\lambda}^{u\,1}_{33})^2+(\tilde{\lambda}^{u\,2}_{33})^2}}
\tilde{t'}_L^c \,, 
\eea 
and the exotic top Yukawa coupling is given
by 
\be \tilde{\lambda}^u_{33} =
\frac{\tilde{\lambda}^{u\,1}_{33}\tilde{\lambda}^{u\,2}_{33}}{\sqrt{(\tilde{\lambda}^{u\,1}_{33})^2+
(\tilde{\lambda}^{u\,2}_{33})^2}} \,. 
\ee 

We can neglect the
mixing between the top, the exotic quark and the components of the
triplets $\tilde{Q}^3_L$ since they are much heavier thanks to the
explicit mass term in \eq{top}.
Therefore, \eq{top} becomes 
\bea 
\label{topd} 
\mathcal{L}^Y_{top}
&=&\lambda^{u}_{33}\,v_W\,t^{c}_L  t_L\big(-k \big) - {\lambda'}^{u}_{33}\,v_F
t^{c}_L \tilde{t}_L
\big(-k \big)+ \hat{m}\,\hat{t}^{c}_L T_L  \big(-k \big) \nn \\
&+& \tilde{\lambda}^{u}_{33}\,v_F\,{\tilde{t}}^c_L \tilde{t}_L
+\tilde{\hat{m}}\,\tilde{\hat{t}}^{c}_L \tilde{T}_L \,,
\eea
where we have neglected the terms involving the exotic
triplets $\tilde{Q}^3_L$. In \eq{topd} there is  mixing
between the standard top $t_L$ and the exotic one $\tilde{t}_L$ which is however very much suppressed, as we shall show shortly.

The same argument should in principle be applied to the first and
 second family. However in these cases we can neglect altogether  the mixing
 between the ${u'}^{{1,2}^{c}}_L$ and ${u''}^{{1,2}^{c}}_L$  because it
 is strongly suppressed. Considering for example the the second family, we  have
\bea
\label{charm}
 c_L^c &= &
\frac{\lambda^{u\,1}_{22}(-\epsilon^2)^3}{\sqrt{(\lambda^{u\,1}_{22}(-\epsilon^2)^3)^2+
(\lambda^{u\,2}_{22})^2}} {c''}^c_L -
\frac{\lambda^{u\,2}_{22}}{\sqrt{(\lambda^{u\,1}_{22}(-\epsilon^2)^3)^2+
(\lambda^{u\,2}_{22})^2}} {c'}_L^c \nn\\
\hat{c}_L^c &= &
\frac{\lambda^{u\,2}_{22}(-\epsilon^2)^3}{\sqrt{(\lambda^{u\,1}_{22}(-\epsilon^2)^3)^2+(\lambda^{u\,2}_{22})^2}}
{c''}_L^c +
\frac{\lambda^{u\,1}_{22}}{\sqrt{(\lambda^{u\,1}_{22}(-\epsilon^2)^3)^2+(\lambda^{u\,2}_{22})^2}}
{c'}_L^c \,, 
\eea
and from \eq{charm} follows that 
\bea
c_L^c &\simeq&  {c'}_L^c \nn\\
\hat{c}_L^c  &\simeq&  {c''}_L^c \,. 
\eea

In the exotic sector the situation follows what happens in the
case of the top quark, and we define a light and a heavy exotic quark for
both families, $\tilde{c}^c_L$ and $\tilde{\hat{c}}^c_L$ for the
second and $\tilde{u}^c_L$ and $\tilde{\hat{u}}^c_L$ for the first
one. At the end, the Yukawa lagrangian for the lightest
quarks, both standard and exotic is given by
\bea
\label{quarkphilight} 
\mathcal{L}^q_Y &=&
\lambda^{u}_{ab}\,v_W\,u^{a^{c}}_L u^b_L \big(-k
\big)^{|y^b_{Q}+ y^a_{\mathcal{U}}|}-{ \lambda'}^{u}_{ab}\,v_F\,
u^{a^{c}}_L \tilde{u}^b_L \big(-k \big)^{|y^b_{Q}+
y^a_{\mathcal{U}}|} \nn \\
&+& \tilde{\lambda}^{u}_{ab}\,v_F \,\tilde{u}_L^{a^{c}}
\tilde{u}_L^b
 \big(-k\big)^{|y^b_{\tilde{U}}- y^a_{\tilde{U}}|}+\lambda^d_{ab} d^{a^{c}}_L d^b_L 2 v_W
\epsilon \big(-k \big)^{|y^b_{Q}+ y^a_{d}|} \,.
 \eea

 To see that the mixing between the standard
 and the exotic fermions is negligible, consider the mass matrix
at the leading order, that is by taking all the parameter $\lambda$
equal to $1$. We have the following $6\times  6$ mass matrix
\bea 
M^u_{RL} &=& \left(
\begin{array}{cccccc}
  v_W k^7 & v_W k^6 &  v_W k^4& v_F k^7 & v_F k^6 &  v_F k^4 \\
  v_W k^5 & v_W k^4& v_W k^2 & v_F k^5 & v_F k^4& v_F k^2 \\
  v_W k^4 & v_W k^3 & v_W k &v_F k^4 & v_F k^3& v_F k \\
  0 & 0 & 0 & v_F & v_F k & v_F k^3 \\
  0 & 0 & 0 & v_F k & v_F &  v_F k^2 \\
  0 & 0 & 0 & v_F k^3 & v_F k^2 & v_F 
 \end{array}\right)   \,.
\eea

To give an estimate of the mixing between standard  and exotic fermions we
have to consider $M^{d^{\dag}}M^d$, that is 
\bea 
\label{mdu}
M^{u^{\dag}}_{RL}M^u_{RL}&=&\left(\begin{array}{cccccc}
  v_W^2 k^8 & v_W^2 k^7& v_W^2 k^5 & v_W v_F  k^8 & v_W v_F k^7& v_W v_F k^5  \\
   v_W^2 k^7& v_W^2 k^6 & v_W^2 k^4 & v_W v_F k^7& v_W v_F k^6 & v_W v_F k^4 \\
  v_W^2 k^5 & v_W^2 k^4 & v_W^2 k^2 &  v_W v_F k^5& v_W v_F k^4 & v_W v_F k^2\\
  v_W v_F  k^8 & v_W v_F k^7& v_W v_F k^5& v_F^2& 2 v_F^2 k& 3 v_F^3  \\
 v_W v_F k^7& v_W v_F k^6 & v_W v_F k^4& 2 v_F^2 & v_F^2 & 2 v_F k^2 \\
  v_W v_F k^5& v_W v_F k^4 & v_W v_F k^2& 3 v_F^3&2 v_F k^2& v_F^2
\end{array}
 \right) \,.
 \eea
Nine angles out of the $15$
 parametrizing the unitary matrix that diagonalize the mass matrix
of \eq{mdu} contribute to the mixing between standard and exotic fermions. Let us call $\theta_{ij}$ with $i=1,2,3$ and $j=4,5,6$ one of these nine angles, we see that 
\be
 \tan 2\theta_{ij}\simeq -2\,
M_{ij}/M_{jj} \,,\ee that is
 \be
\theta_{ij} \simeq - k^{n_{ij}}\, v_W/v_F \,,
\ee
 so the largest mixing angle between the standard and exotic fermions is $\theta_{36} \simeq - k^2\, v_W/v_F
 \simeq \, 10^{-2}$, while that with the first two standard families,
 are completely negligible and  well beyond any current bound~\cite{exotic-bounds}.

\subsection{Leptons}

While standard model quark doublets are put in $SU(3)$ electroweak
antitriplets, standard model left-handed leptons are embedded in
$SU(3)$ triplets. Since leptons are lighter than quarks we should
worry only about the divergent quadratic one-loop corrections to
the flhiggs masses coming from the exotic leptons. The lepton
content of each family is given in table~\ref{tabl}.

\begin{table}[ht]
\begin{center}
\caption{Representations and charges assignments for the leptons. Different families run over the index $i$; they differ only for the flavor charges that are written as $(q_1,q_ 2, q_3)$ for, respectively, the first, second and third family. $U(1)_W$ charges are determined by the data constrains (see text of main body).\label{tabl}}
\vspace{0.2cm}
\begin{tabular}{|c|c|c|c|c|c|}
\hline & &$U(1)_X$\quad&$U(1)_F$\quad&$SU(3)_W$\quad \quad&
$U(1)_W$\quad \\
\hline \hline
$\mathcal{L}^i_L=\left(\begin{array}{c}\tilde{L}^i_L
\\0\\ L_L^i\\ 0\\ \tilde{E}_L^i\\0\\ \end{array}\right)
$ &
\begin{tabular}{c}  $ \tilde{L}^i_L=\left(\begin{array}{c} z^i_L\\ w^i_L\\ \tilde{w}^i_L\\
\end{array} \right)$\\\hline$L^i_L=\left(\begin{array}{c} \nu^i_L\\ e_L^i\\ \tilde{e}_L^i\\
\end{array} \right)$ \\\hline
$\tilde{E}^i_L $  \end{tabular} &
\begin{tabular}{c}\,$\begin{array}{c}\, \\0 \\ \,\\
\end{array}$ \\ \hline $\begin{array}{c} \,\\ 0\\ \,\\
\end{array}$\\ \hline $-1/2$\\ \end {tabular} &
\begin{tabular}{c}$\begin{array}{c}\, \\(-1,1,0)\\ \,\\
\end{array}$ \\ \hline  $\begin{array}{c} \,\\ (-1,1,0)\\ \,\\
\end{array}$ \\ \hline $(-1,1,0)$\\ \end {tabular} & \begin{tabular}{c}$\begin{array}{c}\, \\ \overline{3}\\ \,\\
\end{array}$ \\ \hline $\begin{array}{c} \,\\ 3\\ \,\\
\end{array}$ \\ \hline $1$\\ \end {tabular} &\begin{tabular}{c}$\begin{array}{c}\, \\ -2\\ \,\\
\end{array}$ \\ \hline $\begin{array}{c} \,\\ -4\\ \,\\
\end{array}$ \\ \hline $0$\\ \end {tabular} \\
 \hline $\mathcal{E}^{i^{c}}_L=\left(\begin{array}{c}0
\\e^{c}_L\\0\\ 0\\ 0\\0\\
\end{array}\right)$&${\tilde{e'}}^{i^{c}}_L$ &  $0$&$(-7/2,1/2,3/2)$& $1$
& $3$\\
\hline $\tilde{\mathcal{E}}^{i^{c}}_L=\left(\begin{array}{c}0
\\0\\0\\ 0\\ 0\\{\tilde{e'}}^{c}_L\\
\end{array}\right)$&${\tilde{e'}}^{i c}_L$ &  $1/2$&$(1,-1,0)$& $1$
& $3$\\
\hline $\tilde{L}^{i^{c}}_L$ &$\left(\begin{array}{c} z^{c^{i}}\\
w^{c^{i}}\\ \tilde{w}^{c^{i}}\\ \end{array} \right)$&
 $0$ & $(1, -1, 0)$& $3$&$4$ \\
\hline
&$\tilde{e''}_L^{i c}$& $1/2$& $(1,-1,0)$&$1$&$3$ \\
\hline
\end{tabular}
\end{center}
\end{table}

The right-handed neutrinos $N_i$ with $i=1, \cdots , 8$ couple to
the left-handed triplets. In order to see their effect on the
low-energy lagrangian, it is sufficient to consider a pair of
them, for instance, $\nu^1_R$ and $\tilde \nu^5_R$ since the equal
coupling of the remaining three pairs only renormalizes the
overall Yukawa coupling.

\begin{table}[ht]
\begin{center}
\caption{Representations and charges assignments for the two right-handed
neutrinos.\label{tabnu}} 
 \vspace{0.2cm}
\begin{tabular}{|c|c|c|c|c|}
\hline  &$U(1)_X$\quad&$U(1)_F$\quad&$SU(3)_W$\quad \quad&
$U(1)_W$\quad \\
\hline \hline
$\nu^1_R$&$0$&$1$&$1$&$0$\\
 \hline $\tilde{\nu'}^5_R$&$0$&$-1$&$1$&$0$\\
 \hline $\nu^2_R$&$0$&$0$&$1$&$0$\\
 \hline $\tilde{\nu}^6_R$&$0$&$0$&$1$&$0$\\
 \hline
 \end{tabular}
\end{center}
\end{table}

The Yukawa lagrangian for the leptons is given  at the leading order for each term  by
\bea 
\label{leptonphi} 
\mathcal{L}^l_Y & = & \frac{\eta_1
f}{2}(\overline{\nu_L^{1c}}\tilde{\nu'}_R^5+\overline{\tilde{\nu'}_L^{5c}}\nu_R^1
)+  \frac{\eta_2
f}{2}(\overline{\nu_L^{2c}}\tilde{\nu}_R^6+\overline{\tilde{\nu}_L^{6c}}\nu_R^2
) \nn \\
& +& \lambda^{\nu}_{1a} \,f\, \bar{\nu}_R^1 \Big( \epsilon_{ijk}
L^a_{L_{i}} \Sigma_{4,j} \Sigma_{8,k} \Big)
\Big(\Sigma_{4,4}
\Big)^{|y_{L}^a -1|} + \lambda^{\nu}_{5a}\,f\,\bar{\tilde{\nu'}}_R^5 \Big(
\epsilon_{ijk} L^a_{L_{i}}\Sigma_{4,j} \Sigma_{8,k}  \Big)
\Big(\Sigma_{4,4}\Big)^{|y_{L}^a + 1|} \nn \\
&+& \lambda^{\nu}_{2a} \,f\, \bar{\nu}_R^2 \Big( \epsilon_{ijk}
L^a_{L_{i}} \Sigma_{4,j} \Sigma_{8,k} \Big)
\Big(\Sigma_{4,4}
\Big)^{|y_{L}|} + \lambda^{\nu}_{6a}\,f\,\bar{\tilde{\nu}}_R^6 \Big(
\epsilon_{ijk} L^a_{L_{i}}\Sigma_{4,j} \Sigma_{8,k}  \Big)
\Big(\Sigma_{4,4}\Big)^{|y_{L}^a|} \nn \\
&+& \lambda^{e}_{ab}\,f\, \mathcal{E}^{a^{c}}_L\,\Sigma
\,\mathcal{L}^b_L\Big(\Sigma_{4,4}\Big)^{y_{L}^b -1/2+
y^a_{\mathcal{E}}} \nn\\
&+& \tilde{\lambda}^{e\,1}_{ab}\,f\,
\tilde{\mathcal{E}}^{a^{c}}_L\,\Sigma
\,\mathcal{L}^b_L\Big(\Sigma_{4,4}\Big)^{y_{L}^b -
y^a_{L}} + \tilde{\lambda}^{e\,2}_{ab}\,f\,
{\tilde{e''}}^{a^{c}}_L \tilde{E}_L^b
\Big(\Sigma_{4,4}\Big)^{|y^b_{\tilde{E}}- y^a_{\tilde{E}}|} +
\eta^e_{ab} f \tilde{L}^{a^{c}}_L \tilde{L}^b_L \Big(\Sigma_{4,4}\Big)^{|y^b_{\tilde{L}}- y^a_{\tilde{L}}|}
 \,,
 \eea
with $a=1,2,3$ , $L_L^a$ defined  in \tab{tabl},$\nu_R^{1,2,3}=
\nu^e_R,\nu^\mu_R,\nu^\tau_R$,
$e_L^{{1,2,3}^c}=e_L^c,\mu_L^c,\tau_L^c$,$\tilde{e}_L^{{1,2,3}^c}=\tilde{e}_L^c,\tilde{\mu}_L^c,\tilde{\tau}_L^c$.
$y_{L}^a$ are the flavor charges of the $L_L^a$ and $\tilde{L}_L^a$ triplets members of the multiplets $\mathcal{L}_L^a$ defined in \tabs{tabl}{tabnu},
  $y^a_{\mathcal{E}}$ of the
 $\mathcal{E}_L^{a^{c}}$ multiplets , while the two right handed neutrinos,
 ${\nu}_R^1$ and $\tilde{\nu'}_R^5$, flavor charges are $-1/2$ and $1/2$ respectively. In \eq{leptonphi} we have only two terms that preserve two different subgroups of the approximate global symmetry $SU(10)$, that is the terms with coefficients $\tilde{\lambda}^{e\,(1,2)}$. This permit us to protect the flhiggs $\Phi_{3,4}$ masses from the one-loop quadratic divergent contributions coming from the lepton triplets, since they couple to them with a large Yukawa coupling.
As in \eq{quarkphi} for the quarks, the component $\Sigma_{4,4}$ is the only group singlet of the approximated global symmetries that can be introduced to make the lagrangian flavor invariant.

Like  for the quarks, the exotic leptons $\tilde{e'}_L^{a^{c}}$
 and $\tilde{e''}_L^{a^{c}}$ mix giving a light and a heavy
 exotic leptons, $\tilde{e}_L^{a^{c}}$
 and $\tilde{\hat{e}}_L^{a^{c}}$. In terms of the standard
 leptons, of the light exotic leptons and of the $\Phi_i$,
 \eq{leptonphi} becomes
 \bea
 \label{leptonex}
\mathcal{L}^l_Y & = & \frac{\eta_1
f}{2}(\overline{\nu_L^{1c}}\tilde{\nu'}_R^5+\overline{\tilde{\nu'}_L^{5c}}\nu_R^1
)+  \frac{\eta_2
f}{2}(\overline{\nu_L^{2c}}\tilde{\nu}_R^6+\overline{\tilde{\nu}_L^{6c}}\nu_R^2
)
+\frac{\lambda^{\nu}_{1a}}{f} \bar{\nu}_R^1 \Big( \epsilon_{ijk}
L^a_{L_{i}} \Phi_j \Phi_k\Big)
\Big(\frac{\Phi^\dag_2 \Phi_1}{f^2}
\Big)^{|y_{L}^a -1|}  \nn \\
&+ &\frac{\lambda^{\nu}_{5a}}{f} \bar{\nu}_R^5\Big( \epsilon_{ijk}
L^a_{L_{i}} \Phi_j \Phi_k\Big)
\Big(\frac{\Phi^\dag_2 \Phi_1}{f^2}
\Big)^{|y_{L}^a +1|}+\frac{\lambda^{\nu}_{2a}}{f} \bar{\nu}_R^2 \Big( \epsilon_{ijk}
L^a_{L_{i}} \Phi_j \Phi_k\Big)
\Big(\frac{\Phi^\dag_2 \Phi_1}{f^2}
\Big)^{|y_{L}^a|} +\frac{\lambda^{\nu}_{6a}}{f} \bar{\nu}_R^6 \Big( \epsilon_{ijk}
L^a_{L_{i}} \Phi_j \Phi_k\Big)
\Big(\frac{\Phi^\dag_2 \Phi_1}{f^2}
\Big)^{|y_{L}^a|}  \nn \\
&+& \lambda^{e}_{ab} e^{a^{c}}_L\,\Phi_1^\dag
\,L^b_L\Big(\frac{\Phi_1^\dag \Phi_2}{f^2}\Big)^{y_{L}^b
-1/2+
y^a_{\mathcal{E}}} + \tilde{\lambda}^{e}_{ab} \tilde{e}^{a^{c}}_L\,\Phi^\dag_3
\,L^b_L\Big(\frac{\Phi_1^\dag \Phi_2}{f^2}\Big)^{y_{L}^b
- y^a_{L}} + H.c.
 \eea
 
 The neutrino sector  in \eq{leptoni} is given by four Majorana right-handed neutrinos (two copies of them, actually) and three left-handed neutrinos, these latter being  the standard neutrinos. Right-handed neutrinos are heavy, since their masses is of the same order of the scale $f$, and  we can integrate out them to obtain a Majorana mass matrix for the left-handed ones through the see-saw mechanism~\cite{seesaw}.
If we define the neutrino Dirac mass matrix, $M_{RL_{ia}}^D$ through 
\be
\bar{\nu}_{R\,i} M_{RL_{ia}}^D \nu_L^a  \simeq  \frac{\lambda^{\nu}_{ia}}{f} \bar{\nu}_R^i \Big( \epsilon_{1jk}
L^a_{L_{1}} \Phi_j \Phi_k\Big)
\Big(\frac{\Phi^\dag_2 \Phi_1}{f^2}
\Big)^{|y_{L}^a - y_{R_{\nu}}^i|}  
\ee
and the right-handed Majorana 
mass matrix,   $M_{RR_{ij}}$ by
\be
\overline{\nu_{L}^{i^{c}}}\,M_{RR_{ij}}\,\nu_{R\,j}=  \frac{\eta_1
f}{2}(\overline{\nu_L^{1c}}\tilde{\nu'}_R^5+\overline{\tilde{\nu'}_L^{5c}}\nu_R^1
)+  \frac{\eta_2
f}{2}(\overline{\nu_L^{2c}}\tilde{\nu}_R^6+\overline{\tilde{\nu}_L^{6c}}\nu_R^2
) \,,
\ee
where we have defined $\nu_{R}^T = (\nu_R^1,\tilde{\nu'}_R^5, \nu_R^2,\tilde{\nu}_R^6)$ an $ y_{R_{\nu}}^i$ the right-handed neutrinos flavor charges as reported in \tab{tabnu}, we have
\be
\label{mal}
M_{LL_{ab}}= M^{D^{T}}_{RL_{ia}}M^{-1}_{RR_{ij}}M^D_{RL{jb}} \,. \label{seesaw}
\ee

After the symmetry breakings in \eq{ssb} and after having integrating out the right-handed neutrinos , \eq{leptonex} becomes
\bea
\mathcal{L}^l_Y & = & \tilde{ \lambda}^{\nu}_{ab} \frac{v_W^2}{f} \overline{{\nu}_L^{a^{c}}}  k^{y_L^{l\,a} +y_L^{l\,b} } +  \lambda^{e}_{ab} e_L^{a^{c}}\Big(v_W e^b_L + v_F \tilde{e}^b_L\Big)(-k)^{y_L^b -\frac{1}{2}- y^a_{\mathcal{E}}}+  \tilde{\lambda}^{e}_{ab} \tilde{e}_L^{a^{c}}\tilde{e}^b_L v_F
(k)^{|y_L^b - y^a_{\mathcal{E}}|}  + H.c. \,,
\label{leptoni}
\eea
where we can easily read the left-handed Majorana mass matrix of \eq{mal} and where $O(\tilde{ \lambda}^{\nu}_{ab})= O([\lambda^{\nu}_{ia}]^2)$.

As for the quarks, the mixing between the standard charged leptons and the exotic one is negligible and in the discussion of the textures we will consider only the three standard lepton  families.

The see-saw in \eq{seesaw} is at low energy and therefore provides only a small part of the suppression of the neutrino Yukawa coefficient with respect to the others fermions. The problem of the absolute smallness of  neutrino masses is left unsolved in the flhiggs model which only addresses the relative hierarchy in the fermion masses.

\subsection{Fermion masses and mixing matrices}
\label{sec:mmm}

The fermion mass matrices are obtained from \eq{quarkphilight} and \eq{leptoni}, respectively for quarks and leptons.

The quark mass matrices can be read off from \eq{quarkphilight} by inserting the charges of all fermions according to Table \ref{tabq}. They are given by
\be
{M}^{RL}_{u} =  \lambda^u v_W k
\,\left(
\begin{array}{c c c}
 \lambda^u_{11} k^6 &  \lambda^u_{12} k^5 & \lambda^u_{13}  k^3 \\
 \lambda^u_{21} k^4 &  \lambda^u_{22} k^3  & \lambda^u_{23} k   \\
 \lambda^u_{31} k^3 &  \lambda^u_{32} k^2 &  \lambda^u_{33}
\end{array} \right)
\label{mass-up0}
\ee
and
\be
 {M}^{RL}_{d} =  \lambda^d v_W
k^3 \, \left(
\begin{array}{c c c}
\lambda^d_{11} k^4 &   \lambda^d_{12} k^3 &   \lambda^d_{13}k \\
  \lambda^d_{21}k^3 &   \lambda^d_{22} k^2  &  \lambda^d_{23}   \\
  \lambda^d_{31}k^3 &   \lambda^d_{32} k^2 & \lambda^d_{33}
\end{array} \right) \, ,
\label{mass-down0}
\ee
where, we recall, the texture parameter is given by $k = v_F^2/f^2$. We have written the mass matrices by extracting an overall
coefficient for each matrix according and then treating the ratios of Yukawa
couplings as a set of arbitrary parameters to be varied within a
$O(1)$ range

The essential feature of these mass matrices is that the fundamental textures
are determined by the vacuum structure alone---that is that obtained
by taking all Yukawa couplings $\lambda_{ij}^{u,d}$  of $O(1)$.
In fact, by computing the corresponding CKM matrix
one finds in first approximation
\be
V_{\rm CKM} =
\left( \begin{array}{c c c}
 1 &   O(k) & O(k^3) \\
 O(k) & 1 &   O(k^2) \\
  O(k^3)& O(k^2) &1
\end{array} \right) \, , \label{ckm0}
\ee
that is roughly of the correct form and, moreover,  suggests a value of $k\simeq \sin\theta_C \simeq 0.2$, as anticipated.

At the same time it is possible to extract from (\ref{mass-up0}) and (\ref{mass-down0}) approximated
 mass ratios:
\be
 \frac{m_u}{m_c} \simeq   \frac{m_c}{m_t} \simeq O(k^3) 
 \quad \mbox{and} \quad  
\frac{m_d}{m_s} \simeq \frac{m_s}{m_b}\simeq O(k^2)
\ee
which again roughly agree with the experimental values.

These results show that the quark masses and mixing angles can be reproduced by our textures. While a rough agreement is already obtained by taking alla Yukawa coupling to be equal, the precise agreement with the experimental data depends on the actual choice of the Yukawa couplings $\lambda_{ij}^{u,d}$. Their values can be taken  all of the same order, as we shall see in the appendix, and therefore the naturalness of the model is preserved.

Turning now to the leptons, \eq{leptoni} yields the mass matrices

\be {M}^{LL}_{\nu} =  (\lambda^{\nu})^2 \frac{v^2_W}{\eta_1
f} \,\left(
\begin{array}{c c c}
 {\lambda}^{\nu}_{11} k^2 & {\lambda}^{\nu}_{12}   &{\lambda}^{\nu}_{13} k \\
 {\lambda}^{\nu}_{12}   &{\lambda}^{\nu}_{22} k^2   &{\lambda}^{\nu}_{23} k   \\
{\lambda}^{\nu}_{13} k  &{\lambda}^{\nu}_{23} k   & {\lambda}^{\nu}_{33}
\end{array} \right) \,,
\label{mass-nu}
\ee
where $\lambda^{\nu}$ is  an overall factor of the order of the  yukawa coupling of the neutrino's Dirac mass matrix ,  and
 \be
  {M}^{RL}_{e} =  \lambda^e v_W \, \left(
\begin{array}{c c c}
\lambda^e_{11} k^{3} &   \lambda^e_{12} k^5 &   \lambda^e_{13}k^4 \\
  \lambda^e_{21}k &   \lambda^e_{22} k  &  \lambda^e_{23}  \\
  \lambda^e_{31}k^2 &   \lambda^e_{32}  & \lambda^e_{33} k
\end{array} \right) \, ,
\label{mass-el}
\ee
where again we have extracted the overall factors and written the matrices in terms of the ratios of Yukawa couplings divided by the overall coefficients.

The matrices in
\eqs{mass-nu}{mass-el}  reduce---at the order $O(k)$,
and up to overall factors---to 
\be 
\label{m}
M^{(\nu)} =
\left(\begin{array}{ccc} 0 & 1 & O(k)\cr1 & 0 & O(k)
\cr O(k) & O(k)  & 1
\end{array}\right)
\quad \mbox{and} \quad M^{(l)} =\left(\begin{array}{ccc} 0 &
0 & 0\cr O(k) & O(k) & 1\cr 0&1& O(k)
\end{array}\right)\,, 
\ee
where, as before in the case of the quarks, the 1 stands for $O(1)$ coefficients.

The eigenvalues of $M^{(l)}$ can be computed by diagonalizing $M^{(l)\,\dagger}\,M^{(l)}$.
This product is---again for each entry to leading order in $k$:
\be
M^{(l)\,\dagger}\,M^{(l)}\,=\,\left(
\begin{array}{ccc}
0 & 0 & O(k)\cr 0 & 1 & O(k)\cr O(k) & O(k) & 1
\end{array}
\right)\,.
\label{mm}
\ee

By inspection of the $2\times 2$ sub-blocks, the matrix \eq{mm} is
diagonalized by three rotations with angles, respectively,
$\theta_{23}^l \simeq \pi/4$ and $\theta_{12}^l\simeq \theta_{13}^l \ll 1$,
leading to one maximal mixing angle and two minimal. On the other
hand, the neutrino mass matrix in \eq{m} is diagonalized by three rotations
with angles, rispectively, $\tan 2\theta_{12}^\nu \simeq 2/k^2$ and
$\theta_{23}^\nu\simeq \theta_{13}^\nu \ll 1$
(the label $3$ denotes the heaviest eigenstate).
Therefore, the textures in the mass matrices in \eqs{mass-nu}{mass-el}
give rise to a PMNS mixing matrix~\cite{PMNS}---that is the combination of the
the two rotations above---in which $\theta_{23}$ is maximal,
$\theta_{12}$ is large (up to maximal), while $\theta_{13}$ remains small.

The natural prediction when taking  all  coefficients $O(1)$ is then:
a large atmospheric mixing
angle $\theta_{23}$, possibly maximal, another large solar mixing angle $\theta_{12}$, and a small $\theta_{13}$ mixing angle; at the same time,
the mass spectrum includes one light ($O(k^2)$)
and two heavy states ($O(1)$) in the charged
lepton sector ($m_e$, $m_\mu$ and $m_\tau$ respectively),
two light states ($O(k^2)$) and one heavy ($O(1)$) in the neutrino sector, thus predicting a neutrino spectrum with normal hierarchy.

While the quark textures are the same of those discussed in ref.~\cite{littleflavons}, those for the leptons are slightly different because of the different flavor symmetry, an abelian $U(1)$ in the flhiggs model as opposed to the $SU(2)$ of \cite{littleflavons}.

We have included in the appendix a numerical analysis in which all the experimental data for both  quarks and leptons are reproduced
by a random choice of the rescaled Yukawa coefficients $\lambda_{ij}^{u,d,e,\nu}$ of order 1. This analysis shows that 
we need the texture parameter to be $k=0.14$ and therefore $f\simeq 3.4$ TeV for $v_F \simeq 1.3$ TeV.

\section{Experimental signatures}
\label{sec:es} 

The model contains many new particles. As explained, they are necessary in order to implement the collective symmetry breaking that solve the little hierarchy problem. Some live at the scale $f$, others at the lower scale $v_F$ and  all the way to reach the weak scale $v_W$ below which the standard model particles live. In the low-energy range, these new states affect electroweak precision measurements and, as discussed in sec.~\ref{sec:gbc}, this essentially fixes the scale $v_F$ of  flavor symmetry breaking which cannot be lowered more than about the TeV. They also affect the overall fit of these precision data and can be included together with standard model radiative corrections.

The range of energies from $v_F$ and $f$ is going to be explored in the next few years by LHC. Let us here summarize these new particles predicted by the model and briefly discuss their main experimental signatures.

\begin{table}[ht]
\begin{center}
\caption{Particles and energy spectrum of the model\label{spectrum}}  
\vspace{0.2cm}
\begin{tabular}{|c|c|}
\hline energy scale& states\\
\hline \hline
 $ f \simeq 3$ TeV & $z_{ij},\, s,\, s_{1,2,3},\, W^\prime_{1-8},\, B^\prime,\, X^\prime,\, \nu_R^{1-16},\, \tilde Q_f,\, \tilde L_f$\\
\hline
 $v_F\simeq 1$ TeV &$\tilde W^{\pm}_{1,2},\, Z^\prime,\, X, \, \tilde q_f,\, \tilde l_f,\, \hat q_f,\,\hat l_f,\, \hat{\tilde q}_f,\, \hat{\tilde l}_f$\\
\hline
between  $v_W$  and $v_F$ & $\rho^{\pm}_{1}, \,\delta_{1,3,4},\, \phi_{1-6},\, h^0 ( \phi_7), \, h^\pm ( \rho^{\pm}_{2}) $\\
\hline
 below $v_W = 246$ GeV & $\gamma,\, W^{\pm},\, Z,\, q_f,\, l_f$\\
\hline
\end{tabular}
\end{center}
\end{table}

The most interesting experimental signature for LHC is in the scalar boson sector. The flhiggs model contains 12 scalar bosons, ten of which are neutral, two charged. For arbitrary coefficients of the potential we lack an analytic result for all their masses (see \eqs{mass1}{mass2} for the analytically known part). Their values depend on $v_F$ and $g^\prime$ and, after having fixed them, they are a  function of the parameters of the potential. These parameters $\xi_i$, $\chi_i$ and $\lambda_i$ can assume any  value as long as they remain of order 1. To obtain an estimate of these masses, we vary the numerical value of the coefficients in the potential by a Gaussian distribution around the natural value 1 with a spread of 20\% (that is $\sigma=0.2$). This procedure gives us average values of these masses with a conservative error and we can consider the result the natural prediction of the model. The error is large enough to cover the uncertainty due to higher loop corrections.

 For each solution we verify that all bounds on flavor changing neutral currents are satisfied. The most stringent of these is the potential contribution of the flhiggs fields to the $K^0$-$\bar K^0$ $\Delta S=2$ amplitude. The presence of theflavor-charged  flhiggs fields  at such a low energy scale is possible because the  relevant effective operators induced by their exchange  are suppressed by powers of the fermion masses over $f$~\cite{littleflavons,federica}.

The lightest neutral scalar boson (what would be called the Higgs boson in the standard model) turns out to have a mass 
\be
m_{h^0} = 317 \pm 80 \quad \text{GeV} \, .
\ee
 This is a rather heavy Higgs mass due to the value of $v_F\simeq  1$ TeV we were forced to take in order to satisfy the bounds on the $Z^\prime$ mass. It is still inside the stability bound for a cut off of around a few TeVs. It is a value that  only partially overlaps with the 95\% CL of the overall fit of the  electroweak precison data that gives $m_{h^0} < 237$ GeV~\cite{LEP} and gives the most characteristic prediction of the flhiggs model: a heavy Higgs boson (that is, with a mass larger than 200 GeV). 
 
 Notice that for a heavy Higgs boson like that we  have found, and a cut off $f$ that we  take around 3 TeV in order to generate the correct mass textures, we  would have a little hierarchy problem with a fine tuning of 1\% that justifies the  little-Higgs mechanism we have implemented in order to be solved.

Above the lightest, the other scalar boson masses are spread, the heaviest of them reaching above $f$. The  lightest charged Higgs bosons has a mass $m_{h^\pm} = 560\pm 192$ GeV. 

Like all little-Higgs models, the presence of the heavy gauge bosons and the additional top-like quarks can be used as signatures  in the experimental searches.
In addition, the flhiggs model has also a number of exotic fermionic states of known masses and coupling. They couple only weakly with standard fermions, as explained in section \ref{sec:itf}. They can  be used as further experimental signatures for the model. 

Table \ref{spectrum} lists all the particles present in the flhiggs  model ordered by the energy scale at which they live.

\subsection{Estimating the residual fine tuning}

Even though the model was conceived to provide a framework for electroweak and flavor physics free of fine tuning of the parameters, the requirement of having $v_F\gtap 1$ TeV together with that of having the texture parameter $k$ of the order of the Cabibbo angle---and therefore $f\simeq 3$ TeV---reintroduce some amount of fine tuning. 

The bound on $v_F$ implies relationships on the coefficients of the effective potential that, as already discussed, in turn give a fine tuning of about 10\%.
We find the same amount of fine tuning by considering the effect of having $f\simeq 3$ TeV and therefore of having the exotic quarks related to the top with masses of that order. They give a contribution to the (lightest)  Higgs boson mass of the order of
\be
- \frac{3 f^2 \lambda_t}{16 \pi^2} \log \frac{\Lambda^2}{f^2}
\ee
which, for $m_{h^0} \simeq 300$ GeV is a correction to be cancelled by the bare mass at the 10\% level.

We conclude that while the flhiggs model has still a certain amount of fine tuning in its parameters, this is substantially less than in the standard model with a light Higgs boson.


 \acknowledgments

The authors wish to acknowledge with thanks the
partial support of  the European TMR Networks HPRN-CT-2000-00148
and 00152 and the Benasque Center for Science for hospitality during the completion of this work.

\appendix
\section{Numerical analysis of textures}
\begin{table}[ht]
\label{tabdata}
\begin{center}
\caption{Experimental data vs.\ the result of our numerical analysis based on a representative set of Yukawa couplings of order one (see text)
and  $k=0.14$. Uncertainties in the experimental data are explained in ref.~\cite{littleflavons}}
\label{data}
\vspace{0.2cm}
\begin{tabular}{|c|c|c|}
\hline
\null & exp & numerical results \cr
\hline
\hline
$|V_{us}|$ &   $0.219-0.226$ & 0.22 \cr
$|V_{ub}|$  & $0.002-0.005$ &  0.003\cr
$|V_{cb}|$  & $0.037-0.043$ &   0.04\cr
$|V_{td}|$  & $0.004-0.014$ &  0.007\cr
$|V_{ts}|$  & $0.035-0.043$ &   0.04\cr
$\delta$& $61.5^{o}\pm 7^{o}$  & $53^{o}$ \cr
$\sin 2\beta $ & $0.705^{+0.042}_{-0.032}$ & 0.71\cr
\hline
$m_t/m_c$  & $248\pm 70$ & 222  \cr
$m_c/m_u$ & $325\pm 200$ & 369 \cr
$m_b/m_s$  & $40\pm 10$ &  40 \cr
$m_s/m_d$  & $23\pm 10$ & 17  \cr
\hline
\hline
$\tan^2 \theta_\odot$&$0.23-0.69$ &0.67 \cr
$\sin^2 2\theta_\oplus$ & $0.8-1.0$ & 0.9\cr
$\sin^2 \theta_{13}$ & $<0.09$& 0.03 \cr
\hline
$\Delta m^2_\odot /\Delta m^2_\oplus $&$0.014-0.12$  &0.06 \cr
$m_\tau/m_\mu$& $17$ & 17\cr
$m_\mu/m_e$& $207$ & 190 \\[1ex]
\hline
\end{tabular}
\end{center}
\end{table}

In order to show that the mass textures we found reproduce in a natural manner all the
experimental data we retain the first non-vanishing contribution to
each entry in all mass matrices and then---having extracted an overall
coefficient for each matrix according to \eqs{mass-up0}{mass-down0}
and \eqs{mass-nu}{mass-el}---treat the ratios of Yukawa
couplings as a set of arbitrary parameters to be varied within a
$O(1)$ range. The absolute value of $f$ is immaterial to the textures that only depend on the ratio $k = v_F^2/f^2$. We keep the value of this texture parameter fixed and equal to $k=0.14$. It corresponds in our fit to the values of $v_F = 1260$ GeV and $f=3.4$ TeV.

In practice, we generated for the quark and lepton matrices many sets of Yukawa
parameters whose moduli differ by at most a factor 10
and accepted those that reproduces the known masses and mixings.

For the leptonic sector,  we generate
random sets of 14 real parameters.
Lacking experimental signature
of CP violation in the leptonic sector, we have neglected,
for the purpose of illustration,
leptonic phases in the numerical exercise.

We obtain that for the representative choice
\be
 \left[ \begin{array}{c c c}   {\lambda}^{\nu}_{11}  & {\lambda}^{\nu}_{12}   & {\lambda}^{\nu}_{13}  \\
 {\lambda}^{\nu}_{12}  & {\lambda}^{\nu}_{22}    &{\lambda}^{\nu}_{23}    \\
{\lambda}^{\nu}_{13}   &{\lambda}^{\nu}_{23}    &{\lambda}^{\nu}_{33}
\end{array} \right] =
 \left[ \begin{array}{c c c}
1.6 &  -2.9&  1.0\\
-2.9 & 0.55  &  -0.40 \\
1.0 & -0.40 & 2.9
\end{array} \right]
\ee
with $ \lambda_\nu = O(10^{-5})$ in \eq{mass-nu}, and
\be
\left[ \begin{array}{c c c}  \lambda^e_{11}  &   \lambda^e_{12} &   \lambda^e_{13} \\
  \lambda^e_{21} &   \lambda^e_{22}   &  \lambda^e_{23}   \\
  \lambda^e_{31} &   \lambda^e_{32}  & \lambda^e_{33}
\end{array} \right] =  \left[ \begin{array}{c c c}
 -0.26& -0.83 & 1.0  \\
 -0.48 & -1.7 & -0.13 \\
 -1.2& 2.6 &  -1.1
\end{array} \right]
\ee
with $\lambda^e = O(10^{-2})$ in \eq{mass-el},
the experimental values are well reproduced.

We can see by inspection that there is a certain amount of tension between the request of a maximal mixing angle in the $(2,3)$ sector and the mass splitting between the $\mu$ and $\tau$ that forces an unnatural ratio of about 25 between the smallest and the largest of these ratios of Yukawa coefficients. This was already pointed out in \cite{littleflavons} and is a necessary feature of most textures discussed in the literature

We proceed in a similar manner in the quark sector by generating this time 18 random complex parameters.

We obtain that for the representative choice
\be
\left[ \begin{array}{c c c}  \lambda^u_{11}  &   \lambda^u_{12} &   \lambda^u_{13} \\
  \lambda^u_{21} &   \lambda^u_{22}   &  \lambda^u_{23}   \\
  \lambda^u_{31} &   \lambda^u_{32}  & \lambda^u_{33}
\end{array} \right] =  \left[ \begin{array}{c c c}
 -1.1+1.3 i& 0.37+0.37 i & 0.36 + 0.42 i  \\
 -0.22 -1.6 i & -0.39 - 1.2 i & 1.0 -0.56 i \\
 -0.16 + 1.2 i& 0.39 - 1.1 i &  -1.3 - 0.22 i
\end{array} \right]
\ee
with $\lambda^u = O(k^{-1})$ in \eq{mass-up0}, and
\be
\left[ \begin{array}{c c c}  \lambda^d_{11}  &   \lambda^d_{12} &   \lambda^d_{13} \\
  \lambda^d_{21} &   \lambda^d_{22}   &  \lambda^d_{23}   \\
  \lambda^d_{31} &   \lambda^d_{32}  & \lambda^d_{33}
\end{array} \right] =  \left[ \begin{array}{c c c}
 -0.54 + 1.4 i& -0.38 +0.98 i & -0.85 - 0.09 i  \\
 1.3 - 0.43 i& -0.65 + 0.52 i & 0.51 - 1.2 i \\
 -0.62 - 1.0 i& 0.43 +0.37 i &  -0.02 - 0.54 i
\end{array} \right]
\ee
with $\lambda^d = O(k^{-1})$ in \eq{mass-down0},
the experimental values are well reproduced.

Table \ref{tabdata} summarizes the experimental data and compares them to the result of the above procedure. The agreement is quite impressive.
While the values of the overall constants (which are related to the scale
of the heaviest state in the mass matrices)
are not explained by
the model, the hierarchy among the mass eigenvalues and the
mixing angles are given in first approximation by the flavor symmetry
and the flavor vacuum so that, within each mass matrix,
the Yukawa couplings remain in a natural range.




\begin{thebibliography}{99}




\bibitem{littlest}
N.~Arkani-Hamed, A.~G.~Cohen, E.~Katz and A.~E.~Nelson,
JHEP {\bf 0207}, 034 (2002)
[arXiv:hep-ph/0206021];

\bibitem{skiba}
I.~Low, W.~Skiba and D.~Smith,
Phys.\ Rev.\ D {\bf 66}, 072001 (2002)
[arXiv:hep-ph/0207243];

\bibitem{littlehiggs}
D.~E.~Kaplan and M.~Schmaltz,
JHEP {\bf 0310}, 039 (2003)
[arXiv:hep-ph/0302049].

S.~Chang and J.~G.~Wacker,
Phys.\ Rev.\ D {\bf 69}, 035002 (2004)
[arXiv:hep-ph/0303001].

W.~Skiba and J.~Terning,
Phys.\ Rev.\ D {\bf 68}, 075001 (2003)
[arXiv:hep-ph/0305302].

S.~Chang,
JHEP {\bf 0312}, 057 (2003)
[arXiv:hep-ph/0306034].


\bibitem{federica}
F.~Bazzocchi, Phys.\ Rev.\ D {\bf 70}, 013002 (2004)
[arXiv:hep-ph/0401105].

\bibitem{littlehiggs2}
C.~Csaki, J.~Hubisz, G.~D.~Kribs, P.~Meade and J.~Terning,
Phys.\ Rev.\ D {\bf 67}, 115002 (2003)
[arXiv:hep-ph/0211124];

T.~Han, H.~E.~Logan, B.~McElrath and L.~T.~Wang,
Phys.\ Rev.\ D {\bf 67}, 095004 (2003)
[arXiv:hep-ph/0301040].

\bibitem{Cheng}
H.~C.~Cheng and I.~Low,
arXiv:hep-ph/0405243.


\bibitem{flavormodels-old}
H.~Harari, H.~Haut and J.~Weyers,
Phys.\ Lett.\ B {\bf 78}, 459 (1978).

C.~D.~Froggatt and H.~B.~Nielsen,
Nucl.\ Phys.\ B {\bf 147}, 277 (1979).


T.~Maehara and T.~Yanagida,
Prog.\ Theor.\ Phys.\  {\bf 61}, 1434 (1979).

G.~B.~Gelmini, J.~M.~Gerard, T.~Yanagida and G.~Zoupanos,
Phys.\ Lett.\ B {\bf 135}, 103 (1984).


\bibitem{flavormodels-new} A partial, and by no means complete  list of more recent works on spontaneously broken flavor symmetry includes the following works:


M.~Dine, R.~G.~Leigh and A.~Kagan,
Phys.\ Rev.\ D {\bf 48}, 4269 (1993)
[arXiv:hep-ph/9304299].

M.~Leurer, Y.~Nir and N.~Seiberg,
Nucl.\ Phys.\ B {\bf 398}, 319 (1993)
[arXiv:hep-ph/9212278];
Nucl.\ Phys.\ B {\bf 420}, 468 (1994)
[arXiv:hep-ph/9310320].

P.~Pouliot and N.~Seiberg,
Phys.\ Lett.\ B {\bf 318}, 169 (1993)
[arXiv:hep-ph/9308363].

D.~B.~Kaplan and M.~Schmaltz,
Phys.\ Rev.\ D {\bf 49}, 3741 (1994)
[arXiv:hep-ph/9311281].

L.~J.~Hall and H.~Murayama,
Phys.\ Rev.\ Lett.\  {\bf 75}, 3985 (1995)
[arXiv:hep-ph/9508296].

A.~Pomarol and D.~Tommasini,
Nucl.\ Phys.\ B {\bf 466}, 3 (1996)
[arXiv:hep-ph/9507462];

R.~Barbieri, G.~R.~Dvali and L.~J.~Hall,
Phys.\ Lett.\ B {\bf 377}, 76 (1996)
[arXiv:hep-ph/9512388];

P.~H.~Frampton and O.~C.~Kong,
Phys.\ Rev.\ Lett.\  {\bf 77}, 1699 (1996)
[arXiv:hep-ph/9603372].

E.~Dudas, C.~Grojean, S.~Pokorski and C.~A.~Savoy,
Nucl.\ Phys.\ B {\bf 481}, 85 (1996)
[arXiv:hep-ph/9606383].


P.~Binetruy, S.~Lavignac and P.~Ramond,
Nucl.\ Phys.\ B {\bf 477}, 353 (1996)
[arXiv:hep-ph/9601243].



R.~Barbieri, L.~J.~Hall, S.~Raby and A.~Romanino,
Nucl.\ Phys.\ B {\bf 493}, 3 (1997)
[arXiv:hep-ph/9610449];

G.~Altarelli and F.~Feruglio,
Phys.\ Rept.\  {\bf 320}, 295 (1999).

H.~Fritzsch and Z.~z.~Xing,
Prog.\ Part.\ Nucl.\ Phys.\  {\bf 45}, 1 (2000)
[arXiv:hep-ph/9912358].


Z.~Berezhiani and A.~Rossi,
Nucl.\ Phys.\ B {\bf 594}, 113 (2001)
[arXiv:hep-ph/0003084];

A.~Masiero, M.~Piai, A.~Romanino and L.~Silvestrini,
Phys.\ Rev.\ D {\bf 64}, 075005 (2001)
[arXiv:hep-ph/0104101].


M.~Frigerio and A.~Y.~Smirnov,
Nucl.\ Phys.\ B {\bf 640}, 233 (2002)
[arXiv:hep-ph/0202247].



 \bibitem{littleflavons}
F.~Bazzocchi, S.~Bertolini, M.~Fabbrichesi and M.~Piai,
Phys.\ Rev.\ D {\bf 68}, 096007 (2003)
[arXiv:hep-ph/0306184];
Phys.\ Rev.\ D {\bf 69}, 036002 (2004)
[arXiv:hep-ph/0309182].



\bibitem{bf} F. Bazzocchi and M. Fabbrichesi, {hep-ph/047358}

\bibitem{coleman-weinberg}
S.~R.~Coleman and E.~Weinberg,
Phys.\ Rev.\ D {\bf 7}, 1888 (1973).



\bibitem{FPPS}
M.~Fabbrichesi, R.~Percacci, M.~Piai and M.~Serone,
Phys.\ Rev.\ D {\bf 66}, 105028 (2002)
[arXiv:hep-th/0207013].

\bibitem{seesaw} 
 M. Gell-Mann, P. Ramond, and R. Slansky, in \textit{Supergravity}, Proceedings of the Workshop, Stony Brook, New York, 1979, edited by P. van Nieuwenhuizen and D. Freedman (North-Holland, Amsterdam, 1979); \\
 T. Yanagida, in Proceedings of the Workshop on Unified Theories and Baryon Number in the Universe, Tsukuba, Japan, 1979, edited by A. Sawada and A. Sugamoto (KEK Report No. 79-18, Tsukuba, 1979).

R.~N.~Mohapatra and G.~Senjanovic,
Phys.\ Rev.\ Lett.\  {\bf 44}, 912 (1980).

\bibitem{PMNS} B. Pontecorvo, Sov.\ Phys.\ JETP {\bf 6} (1958) 429;\\
Z. Maki, M. Nakagawa and S. Sakata, Prog.\ Theor.\ Phys.\ {\bf 28} (1962) 870.


\bibitem{PDG}
S. Eidelman et al., Phys.\ Lett.\ {\bf B592} 1, (2004).

\bibitem{Abe}
F.~Abe {\it et al.}  [CDF Collaboration],
Phys.\ Rev.\ Lett.\  {\bf 79}, 2192 (1997).

\bibitem{exotic-bounds}
V.~D.~Barger, M.~S.~Berger and R.~J.~N.~Phillips,
Phys.\ Rev.\ D {\bf 52}, 1663 (1995)
[arXiv:hep-ph/9503204];\\
T.~C.~Andre and J.~L.~Rosner,
Phys.\ Rev.\ D {\bf 69}, 035009 (2004)
[arXiv:hep-ph/0309254].

\bibitem{LEP} The LEP Electroweak Working Group, {\tt http://lepewwg.web.cern.ch/LEPEWWG/plots/winter2004/}.


 \end{thebibliography}
 \end{document}